\documentclass[12pt]{iopart}

\usepackage{iopams}
\usepackage{enumerate}
\usepackage{makeidx}
\usepackage{titletoc}
\usepackage{fancyhdr}
\usepackage{graphicx}% Include figure files
\usepackage{dcolumn}% Align table columns on decimal point
\usepackage{bm}% bold math
\usepackage[dvips]{color}
\usepackage{cite}

\begin{document}

\topical[Electronic conduction properties of indium tin oxide]{Electronic conduction properties of indium tin
oxide: single-particle and many-body transport}

\author{Juhn-Jong Lin$^{1,\ast}$ and Zhi-Qing Li$^2$}
\address{
$^1$NCTU-RIKEN Joint Research Laboratory, Institute of Physics and Department of Electrophysics, National
Chiao Tung University, Hsinchu 30010, Taiwan
\\
$^2$Tianjin Key Laboratory of Low Dimensional Materials Physics and Preparing Technology, Department of
Physics, Tianjin University, Tianjin 300072, China
\\
\vspace*{4mm} $^\ast$Email: jjlin@mail.nctu.edu.tw }

\begin{abstract}

Indium tin oxide (Sn-doped In$_2$O$_{3-\delta}$ or ITO) is an interesting and technologically important
transparent conducting oxide. This class of material has been extensively investigated for decades, with
research efforts focusing on the application aspects. The fundamental issues of the electronic conduction
properties of ITO from 300 K down to low temperatures have rarely been addressed. Studies of the
electrical-transport properties over a wide range of temperature are essential to unraveling the underlying
electronic dynamics and microscopic electronic parameters. In this Topical Review, we show that one can learn
rich physics in ITO material, including the semi-classical Boltzmann transport, the quantum-interference
electron transport, and the electron-electron interaction effects in the presence of disorder and granularity.
To reveal the avenues and opportunities that the ITO material provides for fundamental research, we
demonstrate a variety of charge transport properties in different forms of ITO structures, including
homogeneous polycrystalline films, homogeneous single-crystalline nanowires, and inhomogeneous ultrathin
films. We not only address new physics phenomena that arise in ITO but also illustrate the versatility of the
stable ITO material forms for potential applications. We emphasize that, microscopically, the rich electronic
conduction properties of ITO originate from the inherited free-electron-like energy bandstructure and
low-carrier concentration (as compared with that in typical metals) characteristics of this class of material.
Furthermore, a low carrier concentration leads to slow electron-phonon relaxation, which causes ($i$) a small
residual resistance ratio, ($ii$) a linear electron diffusion thermoelectric power in a wide temperature range
1-300 K, and ($iii$) a weak electron dephasing rate. We focus our discussion on the metallic-like ITO
material.

\end{abstract}

%Uncomment for PACS numbers title message
% Keywords required only for MST, PB, PMB, PM, JOA, JOB?
%Keywords: {indium tin oxide, transparent conducting oxides, electronic conduction properties, Boltzmann transport, localization, mesoscopic physics, granular systems}
%\vspace{2pc}
%\noindent{\it Keywords}: Article preparation, IOP journals
% Uncomment for Submitted to journal title message
%\submitto{\JPCM}
% Comment out if separate title page not required

\maketitle
\pagestyle{plain}

\tableofcontents

\newpage

\section{Introduction}

Transparent conducting oxides (TCOs) constitute an appearing and unique class of materials that simultaneously
possess high electrical conductivity, $\sigma$, and high optical transparency at the visible frequencies
\cite{Holland1955, Jarzebski-PSSa1982,book-Facchetti2010}. These combined electrical and optical properties
render the TCOs to be widely used, for example, as transparent electrodes in numerous optoelectronic devices,
such as flat panel displays, photovoltaic electrochromics, solar cells, energy-efficient windows, and
resistive touch panes \cite{book-Facchetti2010,Ginley-MRS, Granqvist-TSF2002,Granqvis-Solar2007}. Currently,
the major industrial TCO films are made of indium tin oxide (Sn-doped In$_2$O$_{3-\delta}$ or so-called ITO),
F-doped tin oxide, and group III elements doped zinc oxide. Among them, the ITO films are probably the most
widely used TCOs, owing to the ITO's readiness for fabrication and patterning as well as their high quality
and reliability implemented in commercial products.

\begin{figure}
\begin{center}
\includegraphics[scale=1.0]{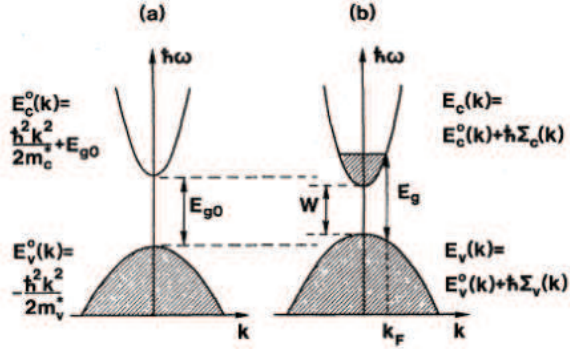}
\caption{Schematic energy bandstructure showing the top of the valence band and the bottom of the conduction band for (a) In$_2$O$_3$ and (b) Sn-doped In$_2$O$_3$ (In$_{2-x}$Sn$_x$O$_{3-\delta}$). This figure was reproduced with permission from \cite{Hamberg-PRB1984}. Copyright 1984 by the American Physical Society.}
\label{banddraftFig}
\end{center}
\end{figure}

On the fundamental research side, our current understanding of the origins for the combined properties of high
electrical conductivity and high optical transparency is based on both theoretical and experimental studies
\cite{Hamberg-PRB1984, Gerfin-JAP1996,Schroer-PRB1993,Imai2003,
Imai2004,Karazhanov-PRB2007,Zunger-PRL2002,Robertson-PRB1984,Mishra-PRB1995,
LZQ-JAP2009,Schleife-PRB2011,LXD-APL2008,Osorio-GuillenPRL2008,Orita-JJAP2010,
Chen-JAP2010,Huy-PRB2011,Yamamoto-PRB2012,Yang-JACS2005,Medevdeva-EPL2005,Odaka-JJPS2001, Mryasov-PRB2001,
Medvedeva-PRL2006, Medvedeva-PRB2010,King-PRL2008,King-PRL2010,Lany-PRL2012,Zhang-PRL2013}. The electronic
energy bandstructure of ITO has been theoretically calculated by several authors \cite{Odaka-JJPS2001,
Mryasov-PRB2001,  Medvedeva-PRL2006, Medvedeva-PRB2010}. It is now known that the bottom of the conduction
band of the parent In$_2$O$_3$ is mainly derived from the hybridization of the In $5s$ electronic states with
the O $2s$ states. The energy-momentum dispersion near the bottom of the conduction band reveals a parabolic
character, manifesting the nature of $s$-like electronic states (see a schematic in
figure~\ref{banddraftFig}). The Fermi level lies in the middle of the conduction and valence bands, rendering
In$_2$O$_3$ a wide-band-gap insulator. Upon doping, the Sn $5s$ electrons contribute significantly to the
electronic states around the bottom of the conduction band, causing the Fermi level to shift upward into the
conduction band. Meanwhile, the shape of the conduction band at the Fermi level faithfully retains the
intrinsic parabolic character. This unique material property makes ITO a highly degenerate n-type
semiconductor or, alternatively, a low-carrier-concentration metal. As a consequence of the $s$-like parabolic
energy bandstructure, the electronic conduction properties of this class of material demonstrate marked
\textit{free-carrier-like} characteristics. The charge transport properties of ITO can thus be quantitatively
described by those simple models formulated basing upon a free electron Fermi gas. Indeed, the levels of close
quantitative agreement between theoretical calculations and experimental measurements obtained for ITO are not
achievable even for alkali (Li, Na, K) and noble (Cu, Ag, Au) metals, as we shall present in this Topical
Review.

In practice, the conduction electron concentration, $n$, in optimally doped ITO (corresponding to
approximately 8 at.\% of Sn doping) can reach a level as high as $n \approx 10^{20}$--$10^{21}$ cm$^{-3}$
\cite{Taga-JVST2000,Rauf-JAP1996}. This level of $n$ is \textit{two to three orders of magnitude lower} than
that ($\approx 10^{22}$--$10^{23}$ cm$^{-3}$ \cite{Kittel}) in typical metals. The room temperature
resistivity can be as low as $\rho$(300\,K) $\approx$ 150 $\mu \Omega$ cm (see table~\ref{TableLi}). This
magnitude is comparable with that of the technologically important titanium-aluminum alloys
\cite{Lin-PRB1993-TiAl,Hsu-PRB1999-TiAl}. In terms of the optical properties, the typical plasma frequency is
$\omega_p \simeq$ 0.7--1 eV \cite{Gerfin-JAP1996}, while the typical energy band gap is $E_g \simeq$ 3.7--4.0
eV. Hence, optimally doped ITO possesses a high optical transparency which exceeds 90\% transmittance at the
visible light frequencies \cite{Guillena-JAP2007,Kim-JAP1999}. A value of $\omega_p \simeq$ 1 eV corresponds
to a radiation frequency of $f_p = \omega_p/2 \pi \simeq 2.4 \times 10^{14}$ Hz, which is approximately one
fifth of the visible light frequency and roughly one fiftieth of the plasma frequency of a typical metal. For
optoelectronic applications, on one hand, one would like to dope ITO with a Sn level as high as
technologically feasible in order to obtain a high electrical conductivity $\sigma$. On the other hand, since
$\omega_p \propto \sqrt{n}$, one has to keep $n$ sufficiently low such that the visible light can propagate
through the ITO structure.

Owing to their technological importance, it is natural that there already exist in the literature a number of
review articles on the ITO as well as TCO materials \cite{Jarzebski-PSSa1982, Ginley-MRS, Chopra,Hamberg-JAP,
Lewis-MRS, Kawazoe-MRS, Minami-MRS, Freeman-MRS, Gordon-MRS, Coutts-MRS, Exarhos-TSF2007, Hosono-TSF2007,
King-JPC2011}. The early studies up to 1982, covering the deposition methods, crystal structures, scattering
mechanisms of conduction electrons, and the optical  properties of In$_2$O$_3$, SnO$_2$ and ITO, were reviewed
by Jarz\c{e}bski \cite{Jarzebski-PSSa1982}. Hamberg and Granqvist discussed the optical properties of ITO
films fabricated by the reactive electron-gun evaporation onto heated glass substrates \cite{Hamberg-JAP}. The
development up to 2000 on the various aspects of utilizing TCOs was summarized in reports considering, for
example, characterizations \cite{Coutts-MRS}, applications and processing \cite{Lewis-MRS}, criteria for
choosing transparent conductors \cite{Gordon-MRS}, new n- and p-type TCOs \cite{Kawazoe-MRS,Minami-MRS}, and
the chemical and thin-film strategies for new TCOs \cite{Freeman-MRS}. The recent progresses in new TCO
materials and TCOs based devices were discussed in \cite{Exarhos-TSF2007} and \cite{Hosono-TSF2007}. King and
Veal recently surveyed the current theoretical understanding of the effects of defects, impurities, and
surface states on the electrical conduction in TCOs \cite{King-JPC2011}.

In this Topical Review, we stress the free-electron-like energy bandstructure and the low-$n$ features (as
compared with typical metals) of the ITO material. These inherited intrinsic electronic characteristics make
ITO a model system which is ideal for not only revealing the semi-classical Boltzmann transport behaviors
(section 2) but also studying new physics such as the quantum-interference weak-localization (WL) effect and
the universal conduction fluctuations (UCFs) in miniature structures (section 3). The responsible electron
dephasing (electron-electron scattering, electron-phonon scattering, and spin-orbit scattering) processes are
discussed. Furthermore, we show that this class of material provides a very useful platform for experimentally
testing the recent theories of \textit{granular metals} \cite{Beloborodov-RMP2007,
Efetov-PRB2003,Efetov-EPL2003,Beloborodov-PRL2003, Kharitonov-PRL2007,Kharitonov-PRB2008}. In the last case,
ultrathin ITO films can be intentionally made to be slightly inhomogeneous or granular, while the coupling
between neighboring grains remains sufficiently strong so that the system retains global metallic-like
conduction (section 4). To illustrate the unique and numerous avenues provided by ITO for the studies of the
aforementioned semi-classical versus quantum electron transport, as well as homogeneous versus inhomogeneous
charge transport, we cover polycrystalline (ultra)thin and thick ITO films and single-crystalline ITO
nanowires in this Topical Review. We demonstrate that high-quality ITO structures can indeed be readily
fabricated into various forms which, apart from being powerful for addressing fundamental electronic
conduction properties, may be useful for potential technological applications. Furthermore, owing to the
similarities in electronic bandstructure between ITO and other TCO materials \cite{book-Facchetti2010}, we
expect that the electronic processes and mechanisms discussed in this Topic Review should be useful for
understanding and interpreting the results obtained on general TCOs.

We do not cover insulating or amorphous ITO materials in this Topical Review, where the electronic conduction
processes can be due to thermally excited hopping \cite{Liu-JAP2003, Iwatsubo-Vacuum2006,
Shigesato-Vacuum2000,LXD-JAP2008,Kytin-APA2014}. In addition to the conventional Mott \cite{Mott-Book1979} and
Efros-Shklovskii \cite{Shklovskii-book1984} hopping conduction mechanisms in homogeneous strongly disordered
systems, electronic conduction due to the thermal charging effect \cite{Sheng-PRL1973} and, more recently, the
variable-range-hopping process \cite{Beloborodov-PRB2005} in inhomogeneous (granular) systems have been
discussed in literature. On the other hand, the possible occurrence of superconductivity in ITO has been
explored in references \cite{Ohyama-JPSJ1985,Mori-JAP1993,Chiu-Nanotechnology2009,Aliev-APL2012}.

\section{Free-electron-like Boltzmann transport: Homogeneous indium tin oxide films and nanowires}

The electrical-transport properties of ITO films have extensively been discussed in the literature. However,
previous studies have mainly concentrated on the influences of deposition methods and conditions on the
$\rho$(300\,K) values. While those studies have provided useful information for improving the fabrication of
high-quality ITO films, they did not deal with the underlying electronic conduction processes in ITO. In
subsection 2.1, we first briefly summarize the theoretical calculations of the electronic energy bandstructure
of ITO and explain why this class of material behaves like a highly degenerate semiconductor or a low-$n$
metal. In subsection 2.2, we discuss the overall temperature behavior of resistivity $\rho(T)$ in ITO and show
that $\rho (T)$ can be well described by the standard Boltzmann transport equation in a wide temperature
range. In subsection 2.3, we demonstrate that the thermoelectric power (Seebeck coefficient, or thermopower),
$S(T)$, in ITO follows an approximately linear temperature dependence in the wide temperature range from 1 K
up to well above room temperature. This \textit{linear thermoelectric power} originates from the diffusion of
electrons in the presence of a temperature gradient and provides a powerful, direct manifestation of the
robust free-carrier-like characteristic of ITO. The reason why the phonon-drag contribution to thermoelectric
power in ITO is absent is heuristically discussed.

\subsection{Free-carrier-like bandstructure and relevant electronic parameters}

\subsubsection{Electronic energy bandstructure}

Since the electronic energy bandstructure plays a key role in governing the charge transport properties of a
given material, we first discuss the electronic bandstructure of ITO. Based on their x-ray photoemission
spectroscopy studies, Fan and Goodenough \cite{Fan-JAP1977} first suggested a schematic energy band model for
the undoped and Sn-doped In$_2$O$_3$ in 1977. A heuristic energy-band model for ITO was proposed by Hamberg
\etal \cite{Hamberg-PRB1984} in 1984. In their heuristic model (shown in figure~\ref{banddraftFig}), the
bottom (top) of the conduction (valence) band of In$_2$O$_3$ was taken to be parabolic. They further proposed
that the shapes of the conduction band and the valence band remained unchanged upon Sn doping. This simple
bandstructure model is qualitatively in line with that obtained by later theoretical calculations
\cite{Odaka-JJPS2001,Mryasov-PRB2001,Medvedeva-PRL2006,Medvedeva-PRB2010}.

\begin{figure}
\begin{center}
\includegraphics[scale=1.2]{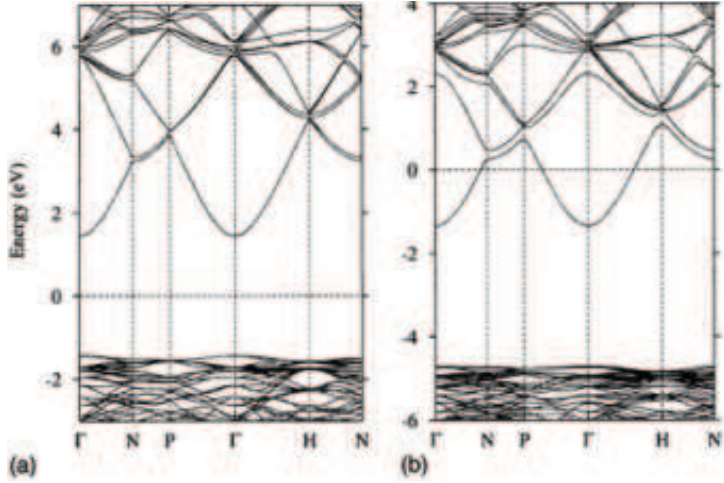}
\caption{Electronic energy bandstructure of (a) undoped In$_2$O$_3$ and (b) 6.25 at.\% Sn-doped In$_2$O$_3$. In (b), a 6.25 at.\% doping level corresponds to a single Sn atom substituting for one of the 16 In atoms in the In$_2$O$_3$ supercell. Note the inherent free-electron-like energy-momentum dispersion at the Fermi level. This figure was reproduced with permission from \cite{Medvedeva-PRB2010}. Copyright 2010 by the American Physical Society.}
\label{Fig-bandMedevedeva}
\end{center}
\end{figure}

The first \emph{ab initio} bandstructure calculations for the ITO material were carried out by Odaka \etal
\cite{Odaka-JJPS2001}, and Mryasov and Freeman \cite{Mryasov-PRB2001} in 2001. Later on, Medvedeva
\cite{Medvedeva-PRL2006} calculated the bandstructure of In$_2$O$_3$, and Medvedeva and Hettiarachchi
\cite{Medvedeva-PRB2010} calculated the bandstructure of 6.25 at.\% Sn-doped In$_2$O$_3$.
Figures~\ref{Fig-bandMedevedeva}(a) and \ref{Fig-bandMedevedeva}(b), respectively, show the electronic
bandstructures of stoichiometric In$_2$O$_3$ and 6.25 at.\% Sn-doped In$_2$O$_3$ obtained in
\cite{Medvedeva-PRB2010}.  For In$_2$O$_3$, the conduction band  exhibits a free-electron-like, parabolic
characteristic around the $\Gamma$ point, where the bottom of the conduction band originates from the
hybridization of In $5s$ and O 2$s$ electronic states. Medvedeva and Hettiarachchi found that the effective
electron mass, $m^\ast$, near the $\Gamma$ point is nearly isotropic. Similar theoretical results were shortly
after obtained by Fuchs and Bechstedt \cite{Fuchs-PRB2008}, and Karazhanov \etal \cite{Karazhanov-PRB2007}.

Upon Sn doping, the Sn $5s$ states further hybridize with the In 5$s$ and O 2$s$ states to form the bottom of
the conduction band. Furthermore, the Fermi level in ITO shifts upward into the conduction band, leading to
the bandstructure depicted in figure~\ref{Fig-bandMedevedeva}(b). Theoretical calculations indicate that the
Sn 5$s$ states contribute nearly one fourth of the total electronic density of states at the Fermi level,
$N(E_F)$, while the In $5s$ and O 2$s$ states contribute the rest. At this particular doping level, the
\emph{s}-like symmetry of the original bandstructure around the Fermi level in the parent In$_2$O$_3$ is
essentially unaltered. Thus, the conduction electrons at the Fermi level in ITO possess strong
free-carrier-like features. Meanwhile, Fuchs and Bechstedt \cite{Fuchs-PRB2008} found that the average
effective electron mass increases slightly with increasing carrier concentration $n$. At a level of $n \simeq
10^{20}$ cm$^{-3}$, they obtained a value $m^\ast \simeq 0.3\,m_e$, where $m_e$ is the free-electron mass.
Their result agreed with that derived from optical measurements of the Drude term to free carriers
\cite{Hamberg-PRB1984}.

In brief, the combined electronic bandstructure characteristics of a wide energy gap, a small $m^\ast$, and in
particular a low $n$ as well as a free-carrier-like dispersion at $E_F$, are the crucial ingredients to make
ITO, on one hand, possess high electrical conductivity while, on the other hand, reveal high optical
transparency.

\subsubsection{Relevant electronic parameters}

\begin{figure}[htp]
\begin{center}
\includegraphics[scale=1.0]{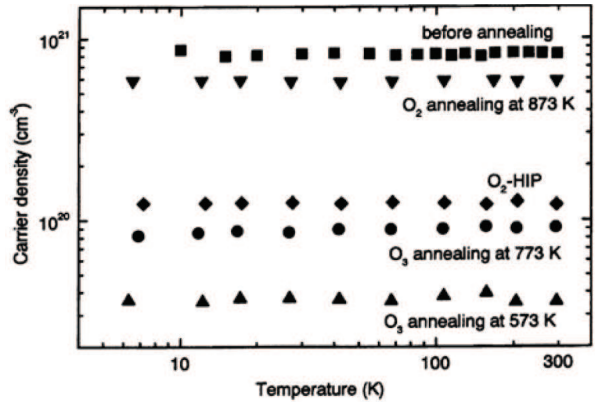}
\caption{Carrier concentration as a function of temperature for as-prepared (before annealing) and annealed ITO films. O$_2$-HIP denotes annealing in an O$_2$ atmosphere, and O$_3$ denotes annealing in an O$_3$ atmosphere. This figure was reproduced with permission from \cite{Kikuchi-Vacuum2000}. Copyright 2000 by the Elsevier B. V.}
\label{Fign-T}
\end{center}
\end{figure}

Experimentally, a reliable method to check the metal-like energy bandstructure of a material is to examine the
temperature $T$ dependence of $n$. For a metal or a highly degenerate semiconductor, $n$ does not vary with
$T$.  Figure~\ref{Fign-T} shows the variation of $n$ with temperature for a few as-deposited (before
annealing) and annealed ITO films studied by Kikuchi \etal \cite{Kikuchi-Vacuum2000}. It is clear that $n$
remains constant in a wide $T$ range from liquid-helium temperatures up to 300 K. In the as-deposited sample,
the $n$ value approaches $\sim 1 \times 10^{21}$ cm$^{-3}$. Temperature independent $n$ in the ITO material
has been reported by a number of groups \cite{LXD-JAP2008,Huang-TSF1987,Shigesato-JAP1993, Taga-JAP1996}.

For the convenience of the discussion of charge transport properties in ITO in this Topical Review, we would
like to estimate the values of relevant electronic parameters. Consider a high-quality ITO sample having a
value of $\rho$(300\,K) $\simeq$ 150 $\mu \Omega$ cm, a carrier concentration $n \simeq 1 \times 10^{21}$
cm$^{-3}$, and an effective mass $m^\ast \simeq$ 0.35\,$m_e$. Applying the free-electron model, we obtain the
Fermi wavenumber $k_F = (3 \pi^2 n)^{1/3} \simeq 3.1 \times 10^9$ m$^{-1}$, the Fermi velocity $v_F = \hbar
k_F/m^\ast \simeq 1.0 \times 10^6$ m/s, and the Fermi energy $E_F = \hbar^2 k_F^2/(2m^\ast) \simeq$ 1.0 eV.
The electron mean free time is $\tau = m^\ast /(ne^2 \rho) \simeq 8.3 \times 10^{-15}$ s, corresponding to the
electron mean free path $l = v_F \tau \simeq$ 8.3 nm. The electron diffusion constant $D = v_F l/3 \simeq$ 28
cm$^2$/s. Thus, the dimensionless product $k_F \l \simeq$ 26. Note that $k_F \l$ is an important physical
quantity which characterizes the degree of disorder in a conductor. A $k_Fl$ value of order a few tens
indicates that high-quality ITO is a weakly disordered metal, and should thus be rich in a variety of
quantum-interference transport phenomena.

In practice, the $\rho$ and $n$ values in ITO films can vary widely with the deposition methods and
conditions, Sn doping levels, and the post thermal treatment conditions.  In table~\ref{TableLi}, we list some
representative values for ITO films prepared by different techniques. This table indicates that those ITO
films fabricated by the DC magnetron sputtering method possess relatively high (low) $n$ ($\rho$) values.
Since the films thus prepared are compact and they adhere well to the substrate surface, this low-cost
technique is thus the most widely used ITO deposition method in the industrial production nowadays. Recently,
researchers have also carried out molecular-beam-epitaxial growth studies of ITO structures
\cite{Tab-2,Tab-3}, but the crystal quality obtained was not as high as that previously achieved in the
epitaxial films grown by a pulsed-laser deposition technique \cite{Ohta-APL2000}. We mention in passing that,
apart from the bulk properties \cite{Nistor-JPCM2010,Seiler-SCM2013}, the effect on electronic processes of
the surface states due to oxygen vacancies in undoped In$_2$O$_{3-\delta}$ \cite{Lany-PRL2012} as well as
doped TCOs \cite{King-PRL2008,King-PRL2010,King-JPC2011} has recently drawn theoretical and experimental
attention.

\begin{table}
\caption{\label{TableLi} Representative values of room temperature resistivity $\rho$ and carrier concentration $n$ for ITO films fabricated by different deposition methods.}
%\begin{ruledtabular}
\begin{center}
\begin{tabular}{lccl}\hline \hline
Fabrication method         &  $\rho$ $(\mu\Omega \, \rm cm)$    &  $n$ ($10^{20}$\,cm$^{-3}$) & References \\  \hline
%Molecular Beam Epitaxy     &    $\lesssim 100$        & $>10$      &  \cite{Tab-2,Tab-3}        \\
Vacuum Evaporation         &  $\sim$ 150--15000    & $\sim$ 5    &  \cite{LXD-JAP2008,Tab-4,Tab-5,Tab-6,Rauf-JAP1996}   \\
Magnetron Sputtering       &    $\sim$ 100--400      & $\sim$ 10   &  \cite{Tab-8,Tab-9,Tab-10,Tab-11}  \\
Chemical Vapor Deposition  &   $\sim$ 150--500      &$\sim$ 10    &  \cite{Tab-12,Tab-13,Tab-14}       \\
Sol-gel                    &  $\sim$  600--4000     & $\sim$ 1   &  \cite{Tab-15,Tab-16,Tab-17,Tab-18}  \\
Spray Pyrolysis            &   $\sim$ 900--5000      & $\sim$ 1  &  \cite{Tab-19,Tab-20,Tab-21}        \\
\hline \hline
\end{tabular}
\end{center}
%\end{ruledtabular}
\end{table}

\subsection{Temperature behavior of electrical resistivity}

\begin{figure}
\begin{center}
\includegraphics[scale=1.0]{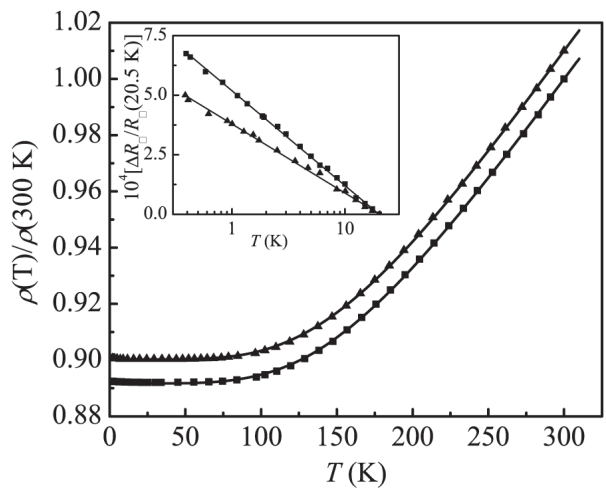}
\caption{Normalized resistivity, $\rho(T)/\rho$(300\,K), as a function of temperature for a 125 nm (squares) and a 240 nm (triangles) thick ITO films. The symbols are the experimental data and the solid curves are the theoretical predictions of equation~(\ref{Bloch-Eq}). For clarity, the data for the 240 nm thick film has been shifted up by 0.01. Inset: Normalized sheet resistance, $\Delta R_\Box (T)/R_\Box = [R_\Box (T) - R_\Box (20.5\,{\rm K})]/R_\Box (20.5\,{\rm K})$, as a function of the logarithm of temperature for these two films below 25 K. The straight solid lines are least-squares fits to the theoretical predictions of 2D WL and EEI effects. This figure was reproduced with permission from \cite{LZQ-JAP2004}. Copyright 2004 by the American Institute of Physics.}
\label{FIGR-T}
\end{center}
\end{figure}

The temperature dependence of resistivity $\rho (T)$ from 300 K down to liquid-helium temperatures provides
key information for the understanding of the electrical conduction processes in a conductor. Li and Lin
\cite{LZQ-JAP2004} have measured $\rho (T)$ between 0.4 and 300 K in a number of 125 and 240 nm thick
polycrystalline ITO films prepared by the standard RF sputtering deposition method. Their films had relatively
low values of $\rho$(300\,K) $\simeq$ 200 $\mu\Omega$ cm. Their results are shown in figure~\ref{FIGR-T}. Li
and Lin found that the $\rho(T)$ data between $\sim$ 25 and 300 K can be well described by the
Bloch-Gr\"{u}neisen formula
\begin{eqnarray} \label{Bloch-Eq}
\rho & = & \rho_e + \rho_{e-{\rm ph}}(T) \nonumber \\ & = & \rho_e + \beta T \left( \frac{T}{\theta_D} \right)^4 \int_0^{\theta_D/T} \frac{x^5{\rm d}x}{(e^x - 1)(1 - e^x)} \,,
\end{eqnarray}
where $\rho_e$ is a residual resistivity, $\beta$ is an electron-phonon ($e$-ph) coupling constant, and
$\theta_D$ is the Debye temperature. The solid curves in the main panel of figure~\ref{FIGR-T} are the
theoretical predications of equation~(\ref{Bloch-Eq}). This figure demonstrates that ITO is a metal, with
$\rho$ decreasing with decreasing temperature (or, a positive temperature coefficient of resistivity, i.e.,
$(1/\rho)(d\rho/dT) > 0$). In particular, the temperature dependence of $\rho (T)$ can be well described by
the standard Boltzmann transport equation.

The first term on the right hand side of equation~(\ref{Bloch-Eq}) originates from the elastic scattering of
electrons with defects. The second term originates from the inelastic scattering of electrons with lattice
vibrations (phonons). Using the Drude formula $\sigma = ne^2 \tau /m^\ast$,  one rewrites $\rho = (m^\ast
/ne^2)(1/\tau_e + 1/\tau_{e-{\rm ph}}) = \rho_e + \rho_{e-{\rm ph}}(T)$, where $e$ is the electronic charge,
$\tau_e$ is the electron elastic mean free time, and $\tau_{e-{\rm ph}}$ is the $e$-ph relaxation time. From
figure~\ref{FIGR-T}, one finds a small resistivity ratio $\rho$(300\,K)/$\rho$(25\,K) $\simeq$ 1.1,
corresponding to the ratio of scattering rates $1/\tau_{e-{\rm ph}} \simeq 0.1(1/\tau_e)$. This observation
explicitly suggests that the $e$-ph relaxation in the ITO material is {\em weak}, and hence the contribution
of the $e$-ph scattering to $\rho$(300\,K) is only approximately one tenth of that of the electron elastic
scattering with imperfections. A slow $e$-ph relaxation rate is a general intrinsic property of low-$n$
conductors, see below for further discussion.\footnote{For comparison, we note that in typical disordered
metals, a measured small residual resistivity ratio $\rho$(300\,K)/$\rho$(4\,K) is usually due to a large
elastic electron scattering rate $1/\tau_e$, because the $e$-ph relaxation is considerably fast in typical
metals, see for example references \cite{Zhong-prl1998,ZhongYL-PRL2010,WuCY-PRB1998}.} The presence a moderate
level of disorder in ITO films result in significant quantum-interference weak-localization (WL) and
electron-electron interaction (EEI) effects at low temperatures. These two effects cause small corrections to
the residual resistivity, which increase with reducing temperature. Close inspection of the inset of
figure~\ref{FIGR-T} indicates a well-defined, logarithmic temperature dependent resistivity rise below $\sim$
25 K. The two-dimensional (2D) WL and EEI effects will be discussed in section 3.

In addition to comparatively thick films, present-day RF sputtering deposition technology has advanced such
that relatively thin films can be made metallic. In a recent study, Lin \etal \cite{Lin-TSF2010} found that
the temperature dependence of $\rho (T)$ below 300 K for 15 nm thick polycrystalline ITO films can also be
described by the Bloch-Gr\"{u}neisen formula. However, the $\rho (T)$ curve reaches a minimum around 150 K. At
lower temperatures, $\rho (T)$ increases with decreasing temperature, signifying much more pronounced 2D WL
and EEI effects than in thicker films (figure~\ref{FIGR-T}).

\begin{figure}
\begin{center}
\includegraphics[scale=0.28]{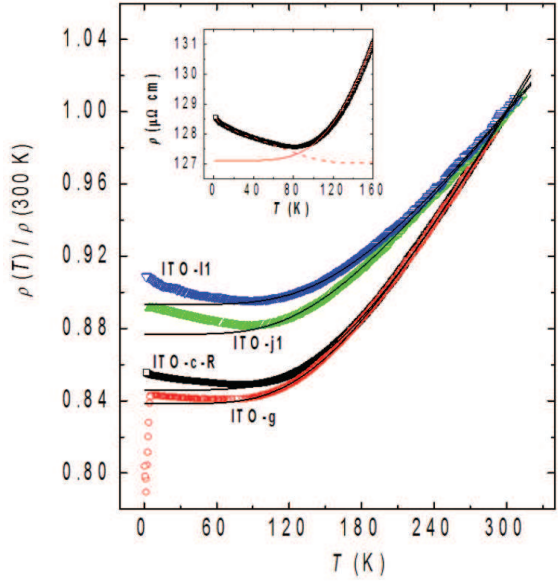}
\caption{Normalized resistivity, $\rho(T)/\rho$(300\,K), as a function of temperature for four single-crystalline ITO nanowires with diameters ranging from 110 to 220 nm. The symbols are the experimental data and the solid curves are the theoretical predictions of equation~(\ref{Bloch-Eq}). At low temperatures, disorder effects cause resistivity rises with reducing temperature. The ITO-g nanowire reveals a possible superconducting transition. The inset shows the measured resistivity as described by the sum of the Bloch-Gr\"{u}neisen law (solid curve) and a disorder-induced correction to the residual resistivity $\rho_0$ (dashed curve) for the ITO-c-R nanowire. This figure was reproduced with permission from \cite{Chiu-Nanotechnology2009}.}
\label{FIGChiuR-T}
\end{center}
\end{figure}

The temperature dependence of resistivity in \textit{single-crystalline} ITO nanowire has been investigated by
Chiu \etal \cite{Chiu-Nanotechnology2009}. They measured individual ITO nanowires from 300 K down to 1.5 K
employing an electron-beam lithographic four-probe configuration. Figure~\ref{FIGChiuR-T} shows a plot of the
normalized resistivity, $\rho(T)/\rho(300\,{\rm K})$, as a function of temperature for four ITO nanowires. The
solid curves are the theoretical predications of equation~(\ref{Bloch-Eq}), indicating that the experimental
$\rho(T)$ data can be well described by the Bloch-Gr\"{u}neisen formula. However, it is surprising that, in
the wide temperature range 1--300 K, the resistivity drops by no more than $\sim$ 20\%, even though these
nanowires are single-crystalline. This observation strongly suggests that these nanowires must contain high
levels of point defects which are not detectable by the high-resolution transmission electron microscopy
studies \cite{Chiu-Nanotechnology2009}. It is worth noting that these nanowires are three-dimensional (3D)
with respect to the Boltzmann transport, because the electron elastic mean free paths $\ell_e = v_F \tau_e
\approx$ 5--11 nm are smaller than the nanowire diameters $d \approx$ 110--220 nm. On the other hand, the
nanowires are one-dimensional (1D) with respect to the WL effect and the UCF phenomena, because the electron
dephasing length $L_\varphi = \sqrt{D \tau_\varphi} > d$ at low temperatures, where $\tau_\varphi$ is the
electron dephasing time (see section 3).

From least-squares fits of the measured $\rho(T)$ to equation~(\ref{Bloch-Eq}), several groups have obtained a
comparatively high Debye temperature of $\theta_D$\,$\sim$\,1000 K in ITO thick films
\cite{LZQ-JAP2004,LXD-JAP2008}, thin films \cite{Lin-TSF2010} and nanowires \cite{Chiu-Nanotechnology2009}.
This magnitude of $\theta_D$ is much higher than those ($\sim$\,200--400 K  \cite{Kittel}) in typical
metals.\footnote{In applying equation~(\ref{Bloch-Eq}) to describe the $\rho (T)$ data in figures~\ref{FIGR-T}
and \ref{FIGChiuR-T}, we have focused on the temperature regime below room temperature. At room temperature
and above, the interaction of electrons with polar optical phonons is strong. By taking into consideration
electron--polar optical phonon interaction, Preissler \etal \cite{Preissler-PRB2013} obtained a value of
$\theta_D \simeq$ 700 K from studies of Hall mobility in In$_2$O$_3$. These studies suggest a high Debye
temperature in the In$_2$O$_3$ based material.}

\begin{figure}
\begin{center}
\includegraphics[scale=1.0]{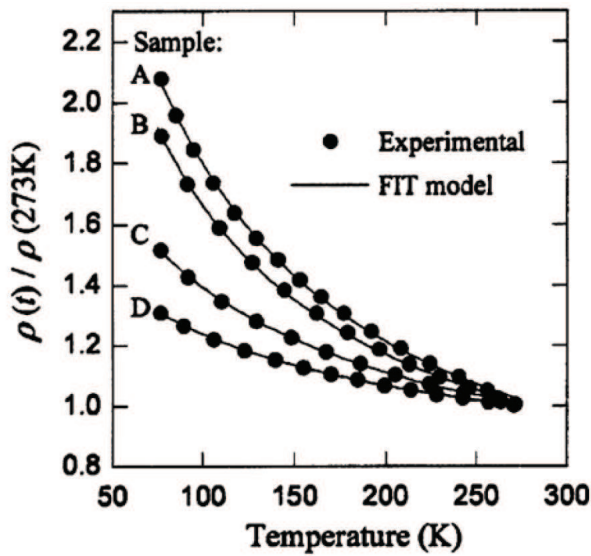}
\caption{Normalized resistivity, $\rho(T)/\rho$(273\,K), as a function of temperature for four ITO nanoparticle films. The nanoparticles have a mean gain size $\approx$ 16 nm, and the films have an approximate thickness $\sim$ 1.1 $\mu$m. The symbols are the experimental data and the solid curves are the theoretical predictions of the thermally fluctuation-induced-tunneling conduction process. This figure was reproduced with permission from \cite{Ederth-PRB2003}. Copyright 2003 by the American Physical Society.}
\label{FIGEderthR-T}
\end{center}
\end{figure}

In addition to films and nanowires, nanoscale ITO particles can be made metallic. Ederth \etal
\cite{Ederth-PRB2003} studied the temperature behavior of porous thin films comprising of ITO nanoparticles.
Their films were produced by spin coating a dispersion of ITO nanoparticles (mean grain size $\approx$ 16 nm)
onto glass substrates, followed by post thermal treatment. They found that the temperature coefficient of
resistivity was negative (i.e., $(1/\rho)(d\rho/dT) < 0$) between 77 and 300 K. However, their $\rho(T)$ data
obeyed the `thermally fluctuation-induced-tunneling conduction' (FITC) process
\cite{Sheng-PRL1978,Sheng-PRB1980,Lin-nanotechnology2008}. Figure~\ref{FIGEderthR-T} shows the normalized
resistivity, $\rho(T)/\rho$(273\,K), as a function of temperature for four ITO nanoparticle films studied by
Ederth \textit{et al}. The symbols are the experimental data, and the solid curves are the FITC theory
predictions. Theoretically, the FITC model considered the global electrical conduction of an inhomogeneous
system consisting of \textit{metal grains} separated by very thin insulating barriers. The thin insulating
barriers were modeled as mesoscopic tunnel junctions. Hence, an observation of the FITC processes occurring in
porous ITO films implies that the constituent ITO nanoparticles are metallic. Indeed, in section 4, we will
discuss that the metallic feature of ITO nanoparticles has provided a powerful platform to experimentally test
the recent theories of granular metals \cite{Beloborodov-RMP2007, Efetov-PRB2003, Efetov-EPL2003,
Beloborodov-PRL2003, Kharitonov-PRL2007, Kharitonov-PRB2008, ZhangYJ-PRB2011}.

We notice in passing that the overall temperature behavior of resistivity in other TCOs, such as Al-doped ZnO
\cite{Bamiduro-APL2007, LiuXD-ASS2012,Yang-APL2012}, Ga-doped ZnO \cite{Bhosle-APL2006,Ahn-APL2007}, Nb-doped
TiO$_2$ \cite{Furubayashi-APL2005, Zheng-ASS2009}, and F-doped SnO$_2$ \cite{Amorim-JPCS2014,Lang-Arxiv1406},
can also be described by the standard Boltzmann transport equation~(\ref{Bloch-Eq}).

\subsection{Linear temperature dependence of thermoelectric power}

The thermoelectric power is an important physical quantity which describes the electronic conduction behaviors
in the presence of a temperature gradient and under the open circuit situation. Studies of the temperature
dependence of thermopower, $S(T)$, can provide useful information about the electronic density of states at
the Fermi level $N(E_F)$, the magnitude of $E_F$, the responsible carrier types (electrons and/or holes), as
well as the phonon-electron and phonon-phonon relaxation processes in the material. In a metal, the
thermopower arises from two contributions and can be expressed as $S(T) = S_d(T) + S_g(T)$, where $S_d (T)$ is
the electron-diffusion contribution, and $S_g (T)$ is the phonon-drag contribution \cite{MacDonald,Siebold}.

\subsubsection{Electron-diffusion thermopower}

The electron diffusion contribution stems from the diffusion of thermal electrons in the presence of a
temperature gradient. A general form is given by the Mott formula \cite{MacDonald}
\begin{equation}\label{Mott_formula}
S_d(T) = - \frac{\pi^2 k_B^2 T}{3|e| E_F} \frac{d \ln \sigma (E)}{d \ln E} \bigg|_{E = E_F} \,,
\end{equation}
where $k_B$ is the Boltzmann constant, and $\sigma (E)$ is the conductivity of electrons that have energy $E$.
The Mott formula is derived under the assumption that the phonon distribution is itself in overall equilibrium
at temperature $T$. Note that in the case of hole conduction the minus sign in equation~(\ref{EqSd-T}) should
be replaced by a plus sign.

Consider a free electron Fermi gas. By substituting the Einstein relation $\sigma (E) = N(E) e^2 D(E)$ into
equation~(\ref{Mott_formula}), where $D(E) = v^2(E) \tau (E)/3$ is the electron diffusion constant in a 3D
conductor with respect to the Boltzmann transport, and $v(E)$ is the electron velocity, one obtains
\begin{equation}\label{Mott_formula_free}
S_d(T) = - \frac{\pi^2 k_B^2 T}{3|e| E_F} \bigg[ \frac32 + \frac{d \ln \tau (E)}{d \ln E} \bigg] \bigg|_{E = E_F} \,.
\end{equation}
Equation~(\ref{Mott_formula_free}) predicts a linear temperature dependence of $S_d$. The slope of this linear
$T$ dependence varies inversely with $E_F$, and its precise value is governed by the energy dependence of
mean-free time $\tau (E) \propto E^q$, where $q$ is an exponent of order unity.

The temperature behavior of $S_d$ in the low temperature limit (which is pertinent to ITO) can be approximated
as follows. At $T \ll \theta_D$ and in the presence of notable defect scattering such that the electron mean
free path $l(E) = v(E) \tau (E)$ is nearly a constant, i.e., $\tau (E) \propto 1/v(E) \propto 1/\sqrt{E}$,
equation~(\ref{Mott_formula_free}) reduces to
\begin{equation}\label{EqSd-T}
S_d = - \frac{\pi^2 k_B^2 T}{3|e|E_F} \,.
\end{equation}
Since the typical $E_F$ value in ITO is one order of magnitude smaller than that in a typical metal, the $S_d$
value in the former is thus approximately one order of magnitude larger than that in the latter.
Alternatively, equation~(\ref{EqSd-T}) can be rewritten in the following form: $S_d = -2C_e/(3n|e|)$, where
$C_e = \pi^2 n k_B^2 T /(2E_F)$ is the electronic specific heat per unit volume. This expression will be used
in equation~(\ref{EqSg-T}).

\begin{figure}
\begin{center}
\includegraphics[scale=1.0]{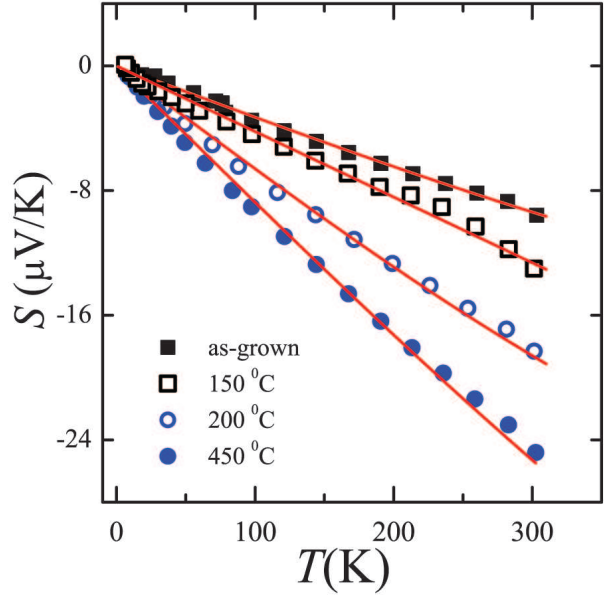}
\caption{Thermoelectric power as a function of temperature for one as-grown and three O$_2$ annealed ITO films. The films were 21 nm thick and the thermal annealing was performed at three different temperatures, as indicated. The straight solid lines are least-squares fits to equation~(\ref{EqSd-T}). This figure was reproduced with permission from \cite{WCY-JAP2010}. Copyright 2010 by the American Institute of Physics.}
\label{FIG-LiS-T}
\end{center}
\end{figure}

The temperature behavior of thermopower in ITO films has been studied by several groups \cite{LZQ-JAP2004,
Lin-TSF2010, WCY-JAP2010, Guilmeau-JAP2009}. Figure~\ref{FIG-LiS-T} shows the measured $S(T)$ data between 5
and 300 K for one as-grown and three annealed ITO films. This figure clearly indicates that $S$ is negative
and varies essentially \textit{linearly} with $T$ in the \textit{wide} temperature range 5--300 K. The
negative sign confirms that electrons are the major charge carriers in ITO.

Recall the fact that the Debye temperature $\theta_D$\,$\sim$\,1000 K in ITO
\cite{LZQ-JAP2004,Chiu-Nanotechnology2009,Preissler-PRB2013}. Therefore, one may safely ascribe the measured
$S$ below 300 K (figure~\ref{FIG-LiS-T}) mainly to the diffusion thermopower $S_d(T)$. The straight solid
lines in figure~\ref{FIG-LiS-T} are least-squares fits to equation~(\ref{EqSd-T}). From the extracted slopes,
one can compute the $E_F$ value in each sample. The value of electron concentration $n$ can thus be deduced
through the free-electron-model expression $E_F = (\hbar^2/2 m^\ast)(3 \pi^2 n)^{2/3}$. In ITO structures, the
extracted values of $E_F$ generally lie in the range $\approx$ 0.5--1 eV  \cite{LZQ-JAP2004,WCY-JAP2010,
Aliev-APL2012}, corresponding to values of $n \approx 10^{20}$--$10^{21}$ cm$^{-3}$. Therefore, ITO can be
treated as a highly degenerate semiconductor or a low-$n$ metal, as mentioned.

\begin{figure}
\begin{center}
\includegraphics[scale=1.0]{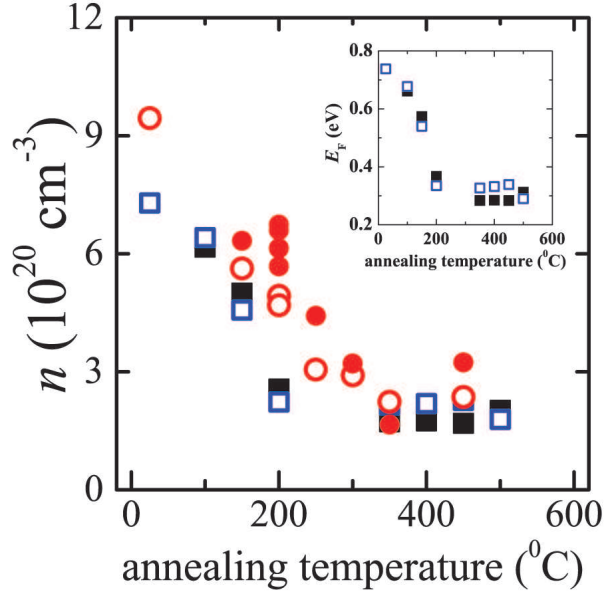}
\caption{Variation in carrier concentration and Fermi energy with annealing temperature for a good number of 21 nm thick ITO films. Open (closed) squares: $n$ for air (oxygen) annealed films; open (closed) circles: $n_H$ for air (oxygen) annealed films. This figure was reproduced with permission from \cite{WCY-JAP2010}. Copyright 2010 by the American Institute of Physics.}
\label{FIG-WuCY}
\end{center}
\end{figure}

It is worth noting that the $n$ values in ITO films obtained from $S(T)$ measurements agree well with those
obtained from the Hall coefficient, $R_H = 1/(n_He)$, measurements. Figure~\ref{FIG-WuCY} shows the extracted
values of $n$ (squares) and the Hall concentration $n_H$ (circles) for a number of as-grown and annealed ITO
films \cite{WCY-JAP2010}. It is seen that the $n$ values agree with the $n_H$ values to within 30\% or better
(except for the films annealed at 200 $^\circ$C, see discussion in \cite{WCY-JAP2010}). This observation
provides a strong experimental support for the validity of the theoretical predictions of a free-carrier-like
energy bandstructure in ITO. In fact, such kind of \textit{prevailing linearity} in $S(T)$ from liquid-helium
temperatures all the way up to at least 300 K (figure~\ref{FIG-LiS-T}) is seldom seen in any textbook simple
metals, where the phonon-drag contribution $S_g(T)$ often causes profound, non-monotonic temperature behavior
of $S(T)$ (see, for example, reference \cite{Siebold} and the figures 7.10 and 7.12 in reference
\cite{Blatt}). Thus, ITO does serve as a model system for studying electronic conduction phenomena and
extracting reliable electronic parameters.

\subsubsection{Phonon-drag thermopower}

We would like to comment on the negligible phonon-drag contribution to the measured $S(T)$ in the ITO
material. The phonon-drag term stems from the interaction between heat conducting phonons with conduction
electrons. In ITO (figures~\ref{FIG-LiS-T}), the prevailing linearity over a wide range of temperature is a
direct and strong indication of the absence of the phonon-drag contribution. The reason for the practically
complete suppression of the phonon-drag term can be explained as follows. Considering the phonon scattering
processes and ignoring their frequency dependence, the phonon-drag thermopower $S_g(T)$ at
$T$\,$<$\,$\theta_D$ can be approximated by \cite{MacDonald,Blatt}
\begin{eqnarray}\label{EqSg-T}
S_g & \simeq & - \frac{C_g}{3n|e|} \left( \frac{\tau_{\rm ph}}{\tau_{\rm ph} + \tau_{{\rm ph}-e}} \right) \nonumber \\ & \simeq & - \frac{C_g}{3n|e|} \left( \frac{\tau_{\rm ph}}{\tau_{{\rm ph}-e}} \right) \simeq \frac12 \left( \frac{\tau_{\rm ph}}{\tau_{e-{\rm ph}}} \right) S_d \,,
\end{eqnarray}
where $C_g$ is the lattice specific heat per unit volume, $\tau_{\rm ph}$ is the phonon relaxation time due to
all kinds of phonon scattering processes (such as phonon-phonon (ph-ph) scattering, phonon scattering with
imperfections, etc.) except the phonon-electron (ph-$e$) scattering, and $\tau_{{\rm ph}-e}$ is the ph-$e$
scattering time. In writing equation~(\ref{EqSg-T}), we have assumed that $\tau_{\rm ph} \ll \tau_{{\rm
ph}-e}$. Note that we have also applied the the energy-balance equation $C_e /\tau_{e-{\rm ph}} = C_g
/\tau_{{\rm ph}-e}$ (references \cite{Reizer1986, Sergeev1996}) to replace $\tau_{{\rm ph}-e}$ by
$\tau_{e-{\rm ph}}$.

Consider a representative temperature of 100 K $\sim$\,0.1$\theta_D$ in ITO. We take the phonon mean free path
to be few nanometers long \cite{Ashida-JAP2009,Cordfunke-JPCS1992,Wang-CI2012}, which corresponds to a
relaxation time $\tau_{\rm ph}$(100\,K)\,$\sim$\,$10^{-12}$ s, with a sound velocity $v_p$\,$\simeq$\,4400 m/s
in ITO \cite{Ashida-JAP2009}. According to our previous studies of the weak-localization effect in ITO films
\cite{Wu-PRB2012}, we estimate $\tau_{e-{\rm ph}}$(100\,K)\,$\sim$\,$10^{-11}$ s. Thus,
equation~(\ref{EqSg-T}) indicates that the phonon-drag term would contribute only a few percent to the
measured thermopower at a temperature of 100 K. The underlying physics for the smallness of the phonon-drag
term $S_g$ can further be reasoned as follows. ($i$) The value of $\tau_{\rm ph}$ in ITO is generally very
short due to the presence of a moderately high level of disorder in this class of material. ($ii$) Since the
$e$-ph coupling strength in a conductor is proportional to the carrier concentration $n$
\cite{Wu-PRB2012,ZhongYL-PRL2010}, the relaxation time $\tau_{e-{\rm ph}}$ in ITO is thus notably long
compared with that in typical metals. (See further discussion in subsection 3.1.2.) These two intrinsic
material characteristics combine to cause a small $\tau_{\rm ph}/\tau_{e-{\rm ph}}$ ratio, and hence $S_g \ll
S_d$ in the ITO material. By the same token, a linear temperature dependence of $S(T)$ with negligible
contribution from $S_g$ has recently been observed in F-doped SnO$_2$ films \cite{Lang-Arxiv1406}.

\section{Quantum-interference transport at low temperature: Homogeneous indium tin oxide films and nanowires}

In section 2, we have examined the temperature dependence of electrical resistivity and thermoelectric power
over a wide temperature range to demonstrate that the electronic conduction properties of metallic ITO obey
the standard Boltzmann transport equation. In particular, being inherited with a free-carrier-like energy
bandstructure, the essential electronic parameters can be reliably extracted from combined $\rho (T)$, $S(T)$
and Hall coefficient $R_H$ measurements. In this section, we show that metallic ITO also opens avenues for the
studies of quantum electron transport properties. We shall focus on the quantum-interference weak-localization
(WL) effect and the universal conductance fluctuation (UCF) phenomenon, which manifest in ITO films and
nanowires at low temperatures. The many-body electron-electron interaction (EEI) effect in homogeneous
disordered systems will not be explicitly discussed in this Topical Review, but will be briefly mentioned
where appropriate.

\subsection{{\label{WL-dephasing}}Weak-localization effect and electron dephasing time}

The WL effect and electron dephasing in disordered conductors have been studied for three decades
\cite{Bergmann-PR1984, Altshuler-EEbook1985, Fukuyama-EEbook1985, Lee-PMP1985,Chakravarty-PR1986,
LinJJ-JPCM2002}. During this time, the mesoscopic and nanoscale physics underlying these processes has
witnessed significant theoretical and experimental advances. Over years, the WL effect has also been explored
in a few TCO materials, including ITO \cite{Wu-PRB2012,Ohyama-JPSJ1983,
Ohyama-JPSJ1985,Chiquito-NL2007,Hsun-PRB2010, YangPY-PRB2012,ZhangYJ-EPL2013}, and ZnO based materials
\cite{LiuXD-JAP2007, Shinozaka-JJSJ2007,Chiu-Nano2013}. In this subsection, we address the experimental 3D,
2D, and 1D WL effects in ITO thick films, thin films, and nanowires, respectively. In particular, we show that
ITO has a relatively long electron dephasing (phase-breaking) length, $L_\varphi(T) = \sqrt{D \tau_\varphi}$,
and a relatively weak $e$-ph relaxation rate $1/\tau_{e-{\rm ph}}$, where $D$ is the electron diffusion
constant, and $\tau_\varphi$ is the electron dephasing time. As a consequence, the WL effect in ITO can
persist up to a high measurement temperature of $\sim$ 100 K. For comparison, in typical normal metals, the WL
effect can often be observed only up to $\sim$ 20--30 K, due to a comparatively strong $e$-ph relaxation rate
as the temperature increases to above liquid-helium temperatures \cite{LinJJ-JPCM2002}. Furthermore, as a
consequence of the small $1/\tau_{e-{\rm ph}}$, one may use ITO thick films to explicitly examine the 3D
small-energy-transfer electron-electron ($e$-$e$) scattering rate, $1/\tau_{ee,{\rm 3D}}^N$, for the first
time in the literature \cite{ZhangYJ-EPL2013}. A long $L_\varphi$ also causes the 1D WL effect and the UCF
phenomenon to significantly manifest in ITO nanowires with diameters $d < L_\varphi$. Since the electronic
parameters, such as $E_F$ and $D$, are well known in ITO, the value of $\tau_\varphi$ can be reliably
extracted and closely compared with the theoretical calculations. Such levels of \textit{close comparison}
between experimental and theoretical values are nontrivial for many typical metals.

\subsubsection{Weak-localization magnetoresistance in various dimensions}

\begin{figure}
\begin{center}
\includegraphics[scale=0.85]{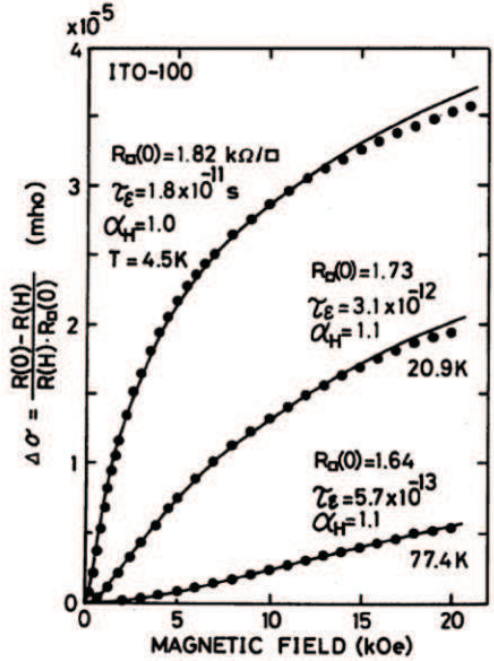}
\caption{Change in the normalized conductivity with the magnetic field for a 7.5 nm thick ITO film at three temperatures, as indicated. The magnetic field was applied perpendicular to the film plane. The solid curves are the predictions of the 2D WL theory. Note that the quantum-interference WL effect persists up to liquid-nitrogen temperatures in the ITO material. This figure was reproduced with permission from \cite{Ohyama-JPSJ1983}. Copyright 1983 by the Physical Society of Japan.}
\label{Ohyama-1983}
\end{center}
\end{figure}

As discussed in section 2, $\rho(T)$ of ITO samples decrease by small amounts ($\lesssim$ 10\% in
polycrystalline films and $\lesssim$ 20\% in single-crystalline nanowires) as the temperature decreases from
300 K down to liquid-helium (or liquid-nitrogen) temperatures, suggesting the presence of moderately high
levels of disorder in all kinds of ITO materials. Thus, the WL effect must prevail in ITO. In 1983, Ohyama
\etal \cite{Ohyama-JPSJ1983} measured ITO thin films and found negative magnetoresistance (MR) and logarithmic
temperature dependence of resistance in a wide temperature range 1.5--100 K. They explained the negative MR in
terms of the 2D WL effect and the logarithmic temperature dependence of resistance in terms of a sum of the 2D
WL and EEI effects. Figure~\ref{Ohyama-1983} shows a plot of the positive magnetoconductance (i.e., negative
MR) induced by the WL effect in a 7.5 nm thick ITO film measured by Ohyama and coworkers. It is seen that the
experimental data (symbols) can be well described by the 2D WL theory predictions (solid curves).

\begin{figure}
\begin{center}
\includegraphics[scale=1.4]{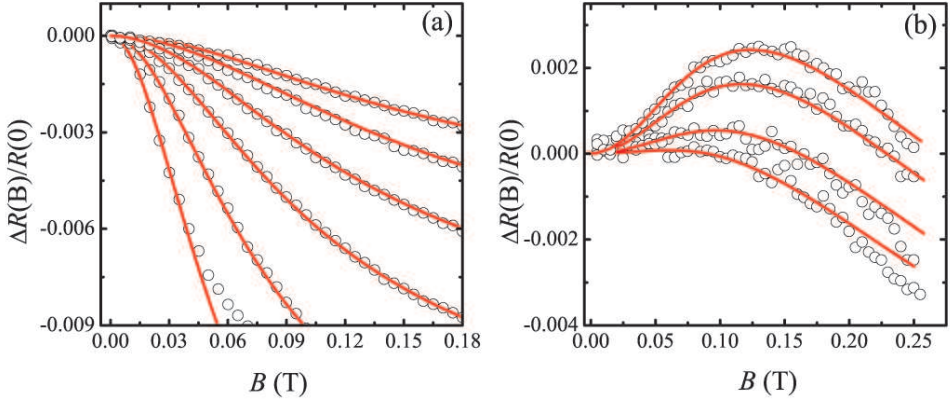}
\caption{Normalized magnetoresistance, $\Delta R(B)/R(0) = [R(B) - R(0)]/R(0)$, as a function of magnetic field of (a) a 60 nm diameter ITO nanowire at (from bottom up): 0.25, 5.0, 12, 20, 30, and 40 K, and (b) a 72 nm diameter ITO nanowire at (from top down): 0.26, 1.0, 2.0, and 4.0 K. The magnetic field was applied perpendicular to the nanowire axis. The symbols are the experimental data and the solid curves are the 1D WL theory predictions. Note that the MRs are negative in (a) (i.e., the weak-localization effect), while positive in (b) (i.e., the weak-antilocalization effect) in small magnetic fields. This figure was reproduced with permission from \cite{Hsun-PRB2010}. Copyright 2010 by the American Physical Society.}
\label{Fig1DMR-H}
\end{center}
\end{figure}

Recently, with the advances of nanoscience and technology, the 1D WL effect has been investigated in
single-crystalline ITO nanowires \cite{Chiquito-NL2007,Hsun-PRB2010, YangPY-PRB2012}. In particular, since
$L_\varphi$ is relatively long in the ITO material at low temperatures (see below), the quasi-1D dimensional
criterion $L_\varphi > d$ is readily achieved. Thus, significant 1D WL effects can be seen in ITO nanowires.
Indeed, figure~\ref{Fig1DMR-H}(a) shows a plot of the negative MR due to the 1D WL effect in a 60 nm diameter
ITO nanowire studied by Hsu \etal \cite{Hsun-PRB2010}. This nanowire had a low resistivity value of
$\rho$(10\,K) $\simeq$ 185 $\mu \Omega$ cm. The magnetic field was applied perpendicular to the nanowire axis.
The data (symbols) is well described by the 1D WL theory predictions (solid curves). The extracted dephasing
lengths are $L_\varphi$(0.25\,K) $\simeq$ 520 nm and $L_\varphi$(40\,K) $\simeq$ 150 nm. Similarly, the
negative MR in the 3D WL effect can be observed in ITO thick films and is well described by the 3D WL theory
predictions. (The explicit theoretical predictions for the 1D, 2D, and 3D MR in the WL effect can be found in
\cite{Chiu-Nano2013} and references therein.)

\subsubsection{Electron dephasing time}

\begin{figure}
\begin{center}
\includegraphics[scale=1.0]{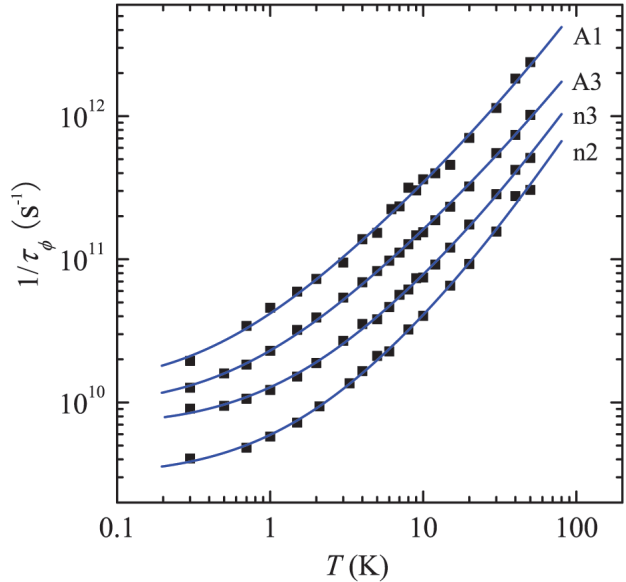}
\caption{Electron dephasing rate $1/\tau_{\varphi}$ as a function of temperature for four 15 nm thick ITO films. The solid curves are least-squares fits to equation~(\ref{Fit2DTao-T}). For clarity, the data for the n3, A3, and A1 films have been shifted up by multiplying by factors of 2, 4, and 8, respectively. This figure was reproduced with permission from \cite{Wu-PRB2012}. Copyright 2012 by the American Physical Society.}
\label{Fig2Dtao-T}
\end{center}
\end{figure}

Measurements of MR in the WL effect allows one to extract the value of $\tau_\varphi$. Detailed studies of the
electron dephasing processes in ITO thin films have recently been carried out by Wu \etal \cite{Wu-PRB2012}.
They have measured the negative MR due to the 2D WL effect and extracted the $\tau_\varphi$ values in two
series of 15 and 21 nm thick ITO films in a wide temperature range 0.3--90 K. Figure~\ref{Fig2Dtao-T} shows a
plot of representative variation of extracted $1/\tau_\varphi$ with temperature. In general, the responsible
dephasing processes are determined by the sample dimensionality, level of disorder, and measurement
temperature \cite{Lee-PMP1985,LinJJ-JPCM2002,Altshuler-JPC1982}. In 3D weakly disordered metals, \emph{e}-ph
scattering is often the dominant dephasing mechanism \cite{LinJJ-JPCM2002,Rammer-PRB1986,Zhong-prl1998}, while
in reduced dimensions (2D and 1D), the \emph{e}-\emph{e} scattering is the major dephasing process
\cite{Altshuler-JPC1982,LinJJ-JPCM2002,Fukuyama-PRB1983,Pierre-PRB2003}. As $T$\,$\rightarrow$\,0 K, a
constant or very weakly temperature dependent dephasing process may exist in a given sample, the physical
origin for which is yet to be fully identified \cite{LinJJ-JPCM2002,
LinJJ-PRB1987b,HuangSM-PRL2007,LinJJ-JPSJ2003, Golubev-PhysicaE2007, Rotter-JPA2009}. In ITO, as already
mentioned, the $e$-ph relaxation rate is very weak.

The total electron dephasing rate $1/\tau_\varphi(T)$ (the solid curves) in figure~\ref{Fig2Dtao-T} for the 2D
ITO thin films studied by Wu \etal \cite{Wu-PRB2012} is described by
\begin{equation}\label{Fit2DTao-T}
\frac{1}{\tau_\varphi (T)} = \frac{1}{\tau_\varphi^0} + A_{ee,{\rm 2D}}^N T + A_{ee,{\rm 2D}}T^2 \ln \left( \frac{E_F}{k_B T} \right) \,,
\end{equation}
where the first, second, and third terms on the right-hand side of the equation stand for the ``saturation"
term, the small-energy-transfer (Nyquist) $e$-$e$ scattering term, and the large-energy-transfer $e$-$e$
scattering term, respectively. The small-energy-transfer term is dominant at low temperatures of $T < \hbar
/(k_B \tau_e)$, while the large-energy-transfer term is dominant at high temperatures of $T > \hbar /(k_B
\tau_e)$. By comparing their measured $1/\tau_\varphi(T)$ with equation~(\ref{Fig2Dtao-T}), Wu \etal found
that their extracted values of the $e$-$e$ scattering strengths $A_{ee,{\rm 2D}}^N \approx 3 \times 10^9$
K$^{-1}$ s$^{-1}$ and $A_{ee,{\rm 2D}} \approx 9 \times 10^6$ K$^{-2}$ s$^{-1}$ are consistent with the
theoretical values to within a factor of $\sim$\,3 and $\sim$\,5, respectively.\footnote{The theoretical
expressions for the small-energy-transfer and large-energy-transfer $e$-$e$ scattering strengths,
respectively, are $A_{ee,{\rm 2D}}^N = (e^2/2 \pi \hbar^2) R_\Box k_B \ln(\pi \hbar /e^2 R_\Box)$ and
$A_{ee,{\rm 2D}} = \pi k_B^2 /(2 \hbar E_F)$, where $R_\Box$ is the sheet resistance. In the comparison of
experiment with theory, the $R_\Box$ value was directly measured, and the $E_F$ value was extracted from
thermoelectric power measurement.} Considering that the ITO material is a disordered
In$_{2-x}$Sn$_x$O$_{3-\delta}$ with random Sn dopants and possible oxygen vacancies, such levels of agreement
between experimental and theoretical values are satisfactory. The good theoretical estimates must derive from
the  free-carrier-like energy bandstructure characteristics of ITO, which renders evaluations of the
electronic parameters reliable. In terms of dephasing length, figure~\ref{Fig2Dtao-T} gives rise to relatively
long length scales of $L_\varphi$(0.3\,K) $\approx$ 500 nm and $L_\varphi$(60\,K) $\approx$ 45 nm.

The $e$-$e$ scattering rate in other-dimensional ITO samples has also been studied. In the case of 1D
nanowires, due to the sample dimensionality effect, the Nyquist $e$-$e$ scattering rate obeys a
$1/\tau_{ee,{\rm 1D}}^N \propto T^{2/3}$ temperature law \cite{LinJJ-JPCM2002,Altshuler-JPC1982}. This
scattering process is largely responsible for the 1D WL MR shown in figures~\ref{Fig1DMR-H}(a) and
\ref{Fig1DMR-H}(b), as analyzed and discussed in \cite{Hsun-PRB2010}. In the case of 3D thick films, the
temperature dependence of the Nyquist rate changes to the $1/\tau_{ee,{\rm 3D}}^N \propto T^{3/2}$ temperature
law \cite{Schmid1974,Altshuler-JETP1979}. Owing to the intrinsic weak $e$-ph coupling in this material, ITO
provides a valuable platform for detailed study of the 3D small-energy-transfer $e$-$e$ scattering process
over wide ranges of temperature and disorder, as discussed below.

In a 3D weakly disordered metal, the $e$-$e$ scattering rate has been calculated by Schmid in 1974 and his
result is given by \cite{Schmid1974}
\begin{equation}\label{Eq.(ee3d)}
\frac{1}{\tau_{ee}} = \frac{\pi}{8} \frac{(k_B T)^2}{\hbar E_F} + \frac{\sqrt{3}}{2\hbar\sqrt{E_F}} \left( \frac{k_B T}{k_F l} \right)^{3/2} \,.
\end{equation}
A similar result has also been obtained by Altshuler and Aronov \cite{Altshuler-JETP1979}. The first term on
the right-hand side of equation~(\ref{Eq.(ee3d)}) is the \emph{e}-\emph{e} scattering rate in a perfect,
periodic potential,  while the second term is the enhanced contribution due to the presence of imperfections
(defects, impurities, interfaces, etc.) in the sample. Microscopically, the second term stands for the Nyquist
\emph{e}-\emph{e} scattering process and is dominant at low temperatures of $T < \hbar/(k_B \tau_e)$, while
the first term represents the large-energy-transfer process and dominates at high temperatures of $T >
\hbar/(k_B \tau_e)$ (references \cite{Altshuler-EEbook1985,Altshuler-JETP1979}). We shall denote the second
term by $1/\tau_{ee,{\rm 3D}}^N = A_{ee,{\rm 3D}}^N T^{3/2}$.  In 3D weakly disordered typical metals, the
\emph{e}-ph scattering is strong and dominates over the $e$-$e$ scattering \cite{LinJJ-JPCM2002}. Thus,
equation~(\ref{Eq.(ee3d)}) has been difficult to test in a quantitative manner for decades, even though the
mesoscopic physics has witnessed marvelous advances.

\begin{figure}[htp]
\begin{center}
\includegraphics[scale=1.4]{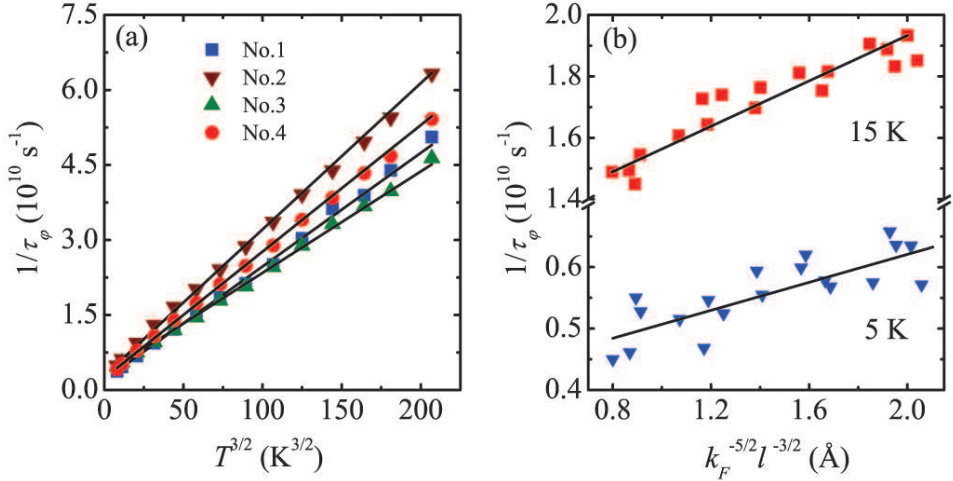}
\caption{(a) Electron dephasing rate $1/\tau_{\varphi}$ as a function of temperature for four ITO thick films. Note that $1/\tau_\varphi$ varies with a $T^{3/2}$ temperature dependence. (b) Variation of $1/\tau_\varphi$ with $k_F^{-5/2}l^{-3/2}$ for a good number of ITO thick films at 5 and 15 K. In the measurement temperature range 4--35 K, the dephasing rate $1/\tau_\varphi \simeq 1/\tau_{ee,{\rm 3D}}^N = A_{ee,{\rm 3D}}^N T^{3/2}$. In (a) and (b), the solid lines are linear fits. This figure was reproduced with permission from \cite{ZhangYJ-EPL2013}.}
\label{Fig3Dtao-T}
\end{center}
\end{figure}

Very recently, Zhang \emph{et al} \cite{ZhangYJ-EPL2013} have measured the low magnetic field MRs in a series
of 3D ITO films with thicknesses exceeding 1 micrometer. Their polycrystalline samples were prepared by the
standard RF sputtering deposition method in an Ar and O$_2$ mixture. During deposition, the oxygen content,
together with the substrate temperature, was varied to ``tune" the electron concentration as well as the
amount of disorder. By comparing the MR data with the 3D WL theory, Zhang \etal extracted the dephasing rate
$1/\tau_{\varphi}$ as plotted in figure~\ref{Fig3Dtao-T}(a). Clearly, one observes a strict $1/\tau_\varphi
\propto T^{3/2}$ temperature dependence in a wide $T$ range 4--35 K. Quantitatively, the scattering rate of
the first term in equation~(\ref{Eq.(ee3d)}) is about one order of magnitude smaller than that of the second
term even at $T$\,=\,35 K in ITO. Thus, the contribution of the first term can be safely ignored. The straight
solid lines in figures~\ref{Fig3Dtao-T}(a) are described by $1/\tau_\varphi = 1/\tau_\varphi^0 + A_{ee,{\rm
3D}}^N T^{3/2}$, where $1/\tau_\varphi^0$ is a constant, and $A_{ee,{\rm 3D}}^N \simeq$ (2.1--2.8)$\times
10^8$ K$^{-3/2}$ s$^{-1}$ for various samples. These experimental $A_{ee,{\rm 3D}}^N$ values are within a
factor of $\sim$\,3 of the theoretical values given by the second term of equation~(\ref{Eq.(ee3d)}).

Furthermore, applying the free-electron model, Zhang \etal \cite{ZhangYJ-EPL2013} rewrote the second term on
the right hand of equation~(\ref{Eq.(ee3d)}) into the form $1/\tau_{ee,{\rm 3D}}^N = A_{ee,{\rm 3D}}^N T^{3/2}
= (1.22 \sqrt{m^\ast} / \hbar^2) (k_B T)^{3/2} k_F^{-5/2}l^{-3/2}$. This expression allows one to check the
combined disorder ($k_F^{-3/2}l^{-3/2}$) and carrier concentration ($k_F^{-1}$) dependence of $1/\tau_{ee,{\rm
3D}}^N$ at a given temperature. Figure~{\ref{Fig3Dtao-T}(b)} shows a plot of the variation of the extracted
$1/\tau_{\varphi}$ with $k_F^{-5/2}l^{-3/2}$ at two $T$ values of 5 and 15 K. Obviously, a variation
$1/\tau_\varphi \propto k_F^{-5/2}l^{-3/2}$ is observed. Quantitatively, the experimental slopes ($\simeq 1.2
\times 10^{19}$ and $3.7 \times 10^{19}$ m$^{-1}$\,s$^{-1}$ at 5 and 15 K, respectively) in
figure~\ref{Fig3Dtao-T}(b) are within a factor of $\sim$\,5 of the theoretical values. Thus, the experimental
dephasing rate $1/\tau_\varphi \simeq 1/\tau_{ee,{\rm 3D}}^N = A_{ee,{\rm 3D}}^N T^{3/2}$ in ITO thick films
quantitatively confirms the temperature, disorder and carrier concentration dependences of the
Schmid-Altshuler-Aronov theory of 3D small-energy-transfer \emph{e}-\emph{e} scattering in disordered metals
\cite{Schmid1974,Altshuler-JETP1979}.

\vspace{6mm}

\textit{Electron-phonon relaxation rate.} We would like to comment on the reason why the \emph{e}-\emph{e}
scattering dominates the electron dephasing rate in 3D ITO thick films (figure~\ref{Fig3Dtao-T}) in a wide $T$
range up to several tens of degrees of kelvin. The reason is owing to the fact that the ITO material possesses
relatively low $n$ values which result in a greatly suppressed $1/\tau_{e-{\rm ph}} \ll 1/\tau_{ee, {\rm
3D}}^N$. Theoretically, it is established that the electron scattering by transverse vibrations of defects and
impurities dominates the \emph{e}-ph relaxation. In the quasi-ballistic limit ($q_Tl>1$, where $q_T$ is the
wavenumber of a thermal phonon),\footnote{In high-quality ITO structures, $q_Tl$\,$\approx$\,0.1\,$T$
\cite{Wu-PRB2012,ZhangYJ-EPL2013}, and hence the quasi-ballistic limit is valid above $\sim$\,10 K. In
disordered normal metals, due to a relatively short electron mean free path $l$\,=\,3$\pi^2 \hbar /(e^2 k_F^2
\rho)$\,$\propto$\,$1/k_F^2$ for a same $\rho$ value, the quasi-ballistic regime is more difficult to realize
in experiment. For example, a polycrystalline Ti$_{73}$Al$_{27}$ alloy \cite{HsuSY-PRB1999} (an amorphous
CuZrAl alloy \cite{LiL-PRB2006}) with $\rho \approx$ 225 $\mu \Omega$ cm ($\approx$ 200 $\mu \Omega$ cm) has a
value of $q_T l \approx 0.006\,T$ ($\approx 0.01\,T$).} the electron-transverse phonon scattering rate is
given by \cite{ZhongYL-PRL2010,Rammer-PRB1986,Sergeev-PRB2000}
\begin{equation}\label{Eq-ep}
\frac{1}{\tau_{e - t,\rm{ph}}}=\frac{3\pi^2 k_B^2 \beta_t}{(p_F u_t)(p_F l)}T^2 \,,
\end{equation}
where  $\beta_t = (2E_F/3)^2N(E_F)/(2\rho_m u_t^2)$ is the electron-transverse phonon coupling constant,
$p_F$ is the Fermi momentum, $u_t$  is the transverse sound velocity, and $\rho_m$ is the mass density. Since
the electronic parameters $E_F$, $p_F$, $N(E_F)$ and $l$ in ITO samples are known, the theoretical value of
equation~(\ref{Eq-ep}) can be computed and is of the magnitude $1/\tau_{e - t,\rm{ph}} \sim$
4$\times$$10^6\,T^2$ K$^{-2}$\,s$^{-1}$. Note that this relaxation rate is about one order of magnitude
smaller than $1/\tau_{ee,{\rm 3D}}^N$ even at a relatively high temperature of 40 K. A weak $e$-ph relaxation
rate allows the quantum-interference WL effect and UCF phenomena to persist up to a few tens of degrees of
kelvin in ITO.\footnote{The electron dephasing length $L_\varphi = \sqrt{D \tau_\varphi} \simeq \sqrt{D
\tau_{e-{\rm ph}}}$ above a few degrees of kelvin is much shorter in a typical disordered metal than in ITO,
due to both a much shorter $\tau_{e-{\rm ph}}$ and a smaller diffusion constant $D \propto 1/(N(E_F) \rho)
\propto 1/N(E_F)$ for a same $\rho$ value in the former.}

We reiterate that equation~(\ref{Eq-ep}) predicts a relaxation rate $1/\tau_{e - t,\rm{ph}} \propto n$. On the
other hand, equation~(\ref{Eq.(ee3d)}) predicts a scattering rate $1/\tau_{ee,{\rm 3D}}^N \propto n^{-5/6}$.
Thus, the ratio of these two scattering rates varies approximately inversely with the square of $n$, namely,
$(1/\tau_{ee,{\rm 3D}}^N)/(1/\tau_{e - t,\rm{ph}}) \propto n^{-2}$. Since the $n$ values in ITO samples are
relatively low, the 3D small-energy-transfer \emph{e}-\emph{e} scattering rate can thus be enhanced over the
\emph{e}-ph relaxation rate. This observation can be extended to other TCO materials, and is worth of further
investigations.

We also would like to note that, in recent studies of superconducting hot electron bolometers, a \textit{weak}
$e$-ph relaxation rate has been observed in quasi-2D heterostructures containing ultrathin
La$_{2-x}$Sr$_x$CuO$_4$ (LSCO) layers \cite{Wen-APL2013}. LSCO has a $n$ value about two orders of magnitude
lower that in the conventional superconductor NbN, and hence $\tau_{e-{\rm ph}}$(LSCO) is nearly two orders of
magnitude longer than $\tau_{e-{\rm ph}}$(NbN). In short, we remark that slow $e$-ph relaxation is a general
intrinsic property of low-$n$ conductors. Generally speaking, one may keep in mind that the relaxation rate
varies approximately as $1/\tau_{e-{\rm ph}} \propto n$ (references \cite{Sergeev-PRB2000,Wen-APL2013}).

\vspace{6mm}

\textit{Spin-orbit scattering time.} According to the recent measurements on a good number of ITO films
\cite{Wu-PRB2012} and nanowires \cite{Hsun-PRB2010} down to as low as 0.25 K, only \textit{negative} MR was
observed (see, for example, figure~\ref{Fig1DMR-H}(a)). This result suggests that the spin-orbit scattering
rate, $1/\tau_{\rm so}$, is relatively weak in ITO. Even at sub-kelvin temperatures where the inelastic
electron scattering events are scarce, one still obtains $1/\tau_{\rm so} < 1/\tau_{ee}^N$(0.25\,K) in many
ITO samples. In other words, the ITO material possesses an inherent long spin-orbit scattering length $L_{\rm
so} = \sqrt{D \tau_{\rm so}}$. In typical ITO films \cite{Wu-PRB2012}, the extracted length scale is $L_{\rm
so} >$ 500 nm, corresponding to a scattering time $\tau_{\rm so} >$ 250 ps. This $\tau_{\rm so}$ value is one
to two orders of magnitude longer than those in typical metals, such as Ag films \cite{Bergmann-PRB1985} and
Sn-doped Ti$_{73}$Al$_{27}$ alloys \cite{Hsu-PRB1999-TiAl}.

In practice, the strength of spin-orbit coupling in a given metal can be tuned by varying the level of
disorder. In general, the spin-orbit scattering rate can be approximately expressed by $1/\tau_{\rm so}
\propto Z^4/\tau_e \propto \rho$, where $Z$ is the atomic number of the relevant (heavy) scatterer. Indeed, an
enhancement of the spin-orbit scattering rate has been achieved in an ITO nanowire which was intentionally
made to have a high resistivity value of $\rho(10\, {\rm K})=1030 \,\mu\Omega\, {\rm cm}$ \cite{Hsun-PRB2010}.
Hsu \etal then observed positive MR at temperatures $T <$ 4 K in low magnetic fields, see
figure~\ref{Fig1DMR-H}(b). A positive MR is a direct manifestation of the weak-antilocalization effect which
results from the scattering rates $1/\tau_{\rm so} > 1/\tau_{ee,{\rm 1D}}^N$ at $T < 4$\,K. At higher
temperatures, a negative MR was recovered, suggesting that $1/\tau_{\rm so} < 1/\tau_{ee,{\rm 1D}}^N $ at $T >
4$\,K. In this high-$\rho$ ITO nanowire, Hsu \etal obtained a moderate length scale $L_{\rm so} \approx$ 95
nm, corresponding to a scattering time $\tau_{\rm so} \approx$ 15 ps. The capability of tuning the spin-orbit
coupling strength might be useful for the future implementation of nanoscale spintronic devices
\cite{Zutic-RMP2004}. Recently, Shinozaki \etal \cite{Shinozaki-JAP2013} have observed an increasing ratio
$(1/\tau_{\rm so})/(1/\tau^N_{ee,{\rm 3D}})$ with increasing $\rho$ in a series of amorphous indium-zinc-oxide
and indium-(tin,gallium)-zinc-oxide thick films.

\subsection{Universal conductance fluctuations}

Universal conductance fluctuations (UCFs) are a fundamental phenomenon in mesoscopic physics. The UCFs
originate from the quantum interference between electron partial waves that propagate along different
trajectories in a miniature system in which \textit{classical self-averaging} is absent or incomplete
\cite{Lee-PRL1985, Lee-PRB1987, Washburn-AdvPhys1986,Beenakker-PRB1988}. Thus, the shape of the UCF patterns
(called `magneto-fingerprints') is very sensitive to the specific {\em impurity configuration} of a given
sample. The UCFs have previously been experimentally observed in lithographic metal and semiconductor
mesoscopic structures at low temperatures \cite{Washburn-AdvPhys1986,Washburn-RPP1992,Thornton-PRB1987}, where
the electron dephasing length $L_{\varphi}$ is comparable to the sample size. Recently, UCFs have been
observed in new artificial materials, including epitaxial InAs nanowires \cite{Hansen-PRB2005}, lithographic
ferromagnets \cite{Wagner-PRL2006}, carbon nanotubes \cite{Man-PRL2005}, graphene
\cite{Berezovsky-NanoTe2010}, and topological insulators \cite{Checkelsky-PRL2011,Chiu-PRB2013}. These new
observations in artificially synthesized materials have enriched and deepened quantum electron transport
physics.

\begin{figure}
\begin{center}
\includegraphics[scale=0.5]{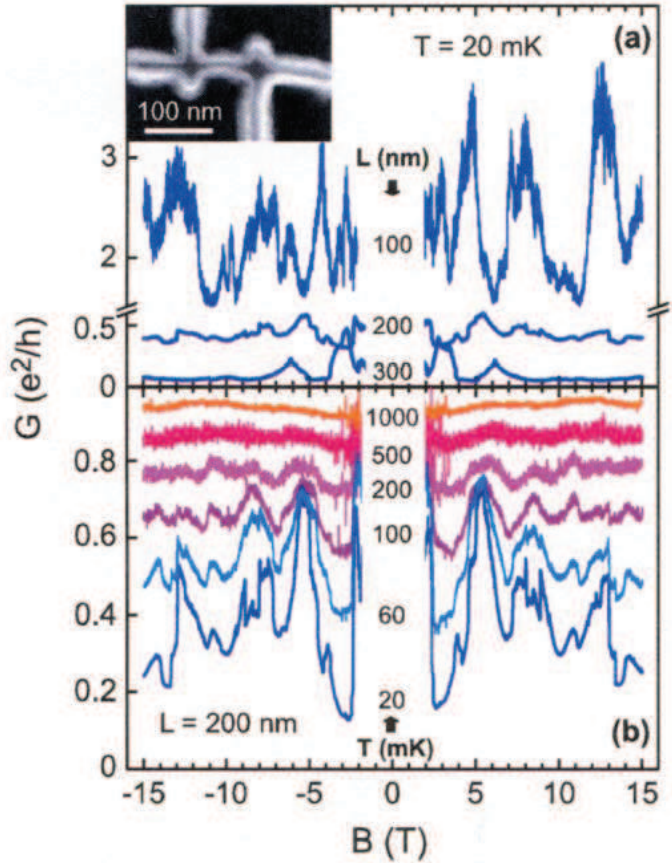}
\caption{(a) Universal conductance fluctuations for three $\sim$\,20 nm wide (Ga,Mn)As wires with different lengths $L \simeq$ 100, 200, and 300 nm. The inset shows an electron micrograph of the 100 nm long wire. (b) Conductance $G$ versus magnetic field $B$ for the 200 nm long wire at several temperatures below 1 K. The magnetic field was applied perpendicular to the wire axis. This figure was reproduced with permission from \cite{Wagner-PRL2006}. Copyright 2006 by the American Physical Society.} \label{Fig-UCF-PRL-2006}
\end{center}
\end{figure}

Wagner \etal \cite{Wagner-PRL2006} have measured the UCFs in lithographically defined ferromagnetic (Ga,Mn)As
nanowires. Figure~\ref{Fig-UCF-PRL-2006}(a) shows their measured conductance $G$ as a function of magnetic
field $B$ for three wires at $T$\,=\,20 mK. The wires were $\sim$\,20 nm wide and 100, 200, or 300 nm long.
Figure~\ref{Fig-UCF-PRL-2006}(b) shows $G$ versus $B$ at several different temperatures between 20 mK and 1 K
for the 200 nm long wire. The magnetic field was applied perpendicular to the wire axis.
Figure~\ref{Fig-UCF-PRL-2006}(b) clearly reveals that the UCFs are observable below $\sim$\,0.5 K.
Figure~\ref{Fig-UCF-PRL-2006}(a) demonstrates that the UCF amplitude significantly decreases with increasing
sample length, suggesting a fairly short dephasing length of $L_\varphi$(20\,mK) $\approx$ 100 nm. For the 100
nm long wire, the peak-to-peak UCF amplitude reaches a value of $e^2/h$ at 20 mK, where $h$ is the Planck
constant.

\vspace{6mm}

\begin{figure}
\begin{center}
\includegraphics[scale=0.5]{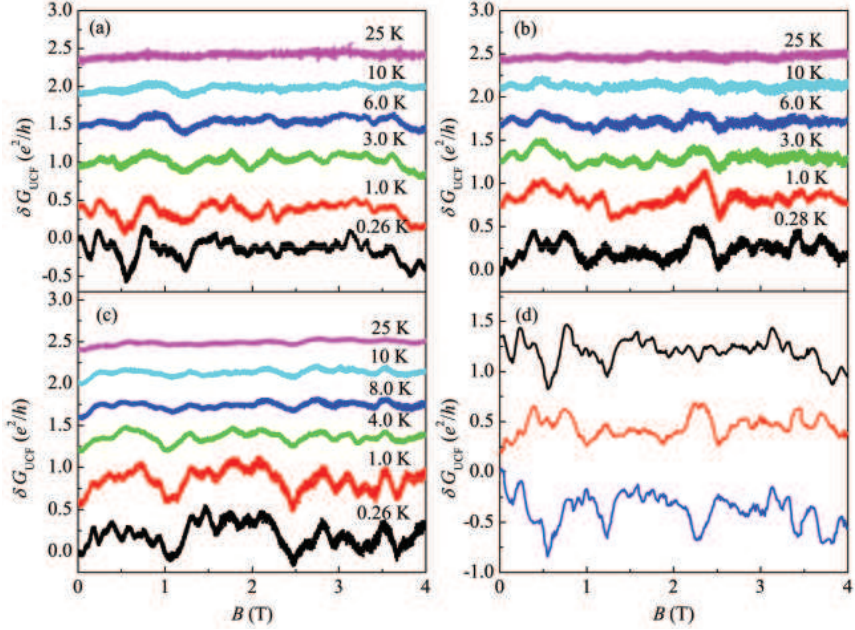}
\caption{Variation of the UCFs, $\delta G_{\rm UCF}(T,B)$, with magnetic field at several temperatures for a 110 nm diameter and 1.2 $\mu$m long ITO nanowire at (a) first cooldown, (b) second cooldown, and (c) third cooldown. (d) The $\delta G_{\rm UCF}(T$\,=\,0.26\,K) curves taken from panel (a) (top curve) and panel (b) (middle curve), and their difference (bottom curve), as a function of magnetic field. The magnetic field was applied perpendicular to the nanowire axis. In panels (a)--(d), the UCF curves are vertically offset for clarity. This figure was reproduced with permission from \cite{YangPY-PRB2012}. Copyright 2012 by the American Physical Society.
\label{Fig1-UCFs}}
\end{center}
\end{figure}

\textit{Impurity reconfiguration.} Let us return to the case of ITO. Since $L_\varphi$ can reach $\approx$ 500
nm at low temperatures, the ITO nanowires are very useful for the investigations of the 1D UCF phenomena. Yang
\etal \cite{YangPY-PRB2012} have recently carried out the magneto-transport measurements on individual ITO
nanowires with a focus on studying the UCFs. Their nanowires were made by implanting Sn ions into
In$_2$O$_{3-\delta}$ nanowires. Figures~\ref{Fig1-UCFs}(a)--(d) show four plots of the variation of the UCFs,
denoted by $\delta G_{\rm UCF}(T,B)$, with magnetic field $B$ for a 110 nm diameter ITO nanowire at several
temperatures.\footnote{The universal conductance fluctuation $\delta G_{\rm UCF}(T,B)$ is defined by
subtracting a smooth magneto-conductance background (including the WL MR contribution) from the measured
$G(T,B)$.} The magnetic field was applied perpendicular to the nanowire axis. Here, after the first run at
liquid-helium temperatures, the nanowire was thermally cycled to room temperature, at which it stayed
overnight, and cooled down again for the magneto-transport measurements at liquid-helium temperatures. The
thermal cycling to room temperature was repeated twice, and the sample was thus measured for three times at
three different cooldowns. The idea was that a thermal cycling to 300 K could possibly induce {\em impurity
reconfiguration} in the given nanowire. A new impurity configuration must lead to differing trajectories of
the propagating electron partial waves, which in turn cause distinct quantum interference. As a result, the
shape of the UCF patterns should be completely changed. Figure~\ref{Fig1-UCFs}(a) shows $\delta G_{\rm
UCF}(T,B)$ as a function of $B$ at several temperatures measured at the first cooldown.
Figure~\ref{Fig1-UCFs}(b) shows $\delta G_{\rm UCF}(T,B)$ as a function of $B$ at several temperatures
measured at the second cooldown, and figure~\ref{Fig1-UCFs}(c) shows those measured at the third cooldown.

A number of important UCF features and the underlying physics can be learned from close inspection of these
figures.
\begin{description}
\item[($i$)] Inspection of figures~\ref{Fig1-UCFs}(a)--(c) indicates that the UCF magnitudes decrease with increasing temperature and disappear at $\sim$\,25 K. Thus, these quantum conductance fluctuations are distinctly different from the classical thermal noise whose resistance fluctuation magnitudes increase with increasing temperature.

\item[($ii$)] During a given cooldown, the shape of the UCF patterns at different temperatures remains the same to a large extent. This observation implies that the impurity configuration is frozen for a considerable period of time if the nanowire is constantly kept at liquid-helium temperatures. A given impurity configuration gives rise to a specific `magneto-fingerprint,' strongly suggesting that the UCF phenomena is a robust manifestation of an intrinsic quantum-interference effect.

\item[($iii$)] At a given temperature, the UCFs among different cooldowns reveal similar peak-to-peak magnitudes.

\item[($iv$)] Figure~\ref{Fig1-UCFs}(d) shows a plot of the $\delta G_{\rm UCF}(T$\,=\,0.26\,K,$B)$ curves taken from figure~\ref{Fig1-UCFs}(a) (top curve) and figure~\ref{Fig1-UCFs}(b) (middle curve), and their difference (bottom curve). This figure is convenient for close inspection and comparison. The top two curves reveal completely different shapes of the UCF patterns, strongly reflecting that a thermal cycling to 300 K has induced an impurity reconfiguration. On the other hand, the UCF magnitudes of these two curves retain similar, with a peak-to-peak value of $\delta G_{\rm UCF}$($T$\,=\,0.26\,K) $\approx 0.5 e^2/h$ for both curves. The reason for retaining a similar UCF magnitude is as follows. The UCF magnitudes in a given nanowire are governed by the $L_\varphi$ values, which are determined by the level of disorder, i.e., the $\rho$ value (or the $R_\Box$ value in 2D), see subsection 3.1.2. The $\rho$ ($R_\Box$) value of a sample is determined by the total number of impurities, but insensitive to the specific spatial distribution of the impurities (provided that the impurity concentration is uniform throughout the sample).\footnote{The UCF studies also allow extractions of the $L_\varphi (T)$ values in a miniature sample. The values thus obtained are in fair accord with those extracted from the WL MR measurements. In addition to $L_\varphi$, the thermal diffusion length $L_T$ plays a key role in governing the UCF magnitudes.}\\
\end{description}

\textit{Classical self-averaging and thermal averaging at finite temperatures.} In the case of a quasi-1D wire
with length $L$, the UCF theory predicts a root-mean-square conductance fluctuation magnitude of
$\sqrt{{\langle (\delta G_{\rm UCF})^2 \rangle}} \simeq 0.73 e^2/h$ in the limit of $T \rightarrow 0$ K
\cite{Lee-PRL1985, Lee-PRB1987, Beenakker-PRB1988}. At this low $T$ limit, the wire behaves as a single
phase-coherent regime. As the temperature gradually increases from absolute zero, $L_\varphi (T)$ becomes
progressively shorter and one has to take into account the classical self-averaging effect. That is, the
phase-coherent regime is expected to be cut off by $L_\varphi$ and the UCF magnitude $\sqrt{{\langle (\delta
G_{\rm UCF})^2 \rangle}}$ is predicted to be suppressed by a factor $(L_\varphi/L)^{3/2}$ under the condition
$L_\varphi < L_T$, where $L_T = \sqrt{D \hbar/k_BT} \propto 1/\sqrt{T}$ is the thermal diffusion length
defined in the EEI theory. The suppression of the UCF magnitudes originates from the fact that the UCFs of
different phase-coherent regimes fluctuate statistically independently. If the temperature further increases
such that $L_T < L_\varphi$ or, equivalently, the thermal energy exceeds the Thouless energy $k_BT >
\hbar/\tau_\varphi$, one also has to take into account the \textit{thermal averaging} effect. That is, the
phase-coherent regime is now expected to be cut off by $L_T$ and the UCF magnitude $\sqrt{{\langle (\delta
G_{\rm UCF})^2 \rangle}}$ is predicted to be suppressed by a factor $(L_T/L) \sqrt{L_\varphi/L}$. These
theoretical concepts have been well accepted by the mesoscopic physics communities for three decades, but have
rarely been experimentally tested in a quantitative manner. The lack of experimental information was mainly
due to the fact that the UCFs could be observed only at temperatures below 1 K in conventional lithographic
metal and semiconductor mesoscopic structures. Fortunately, the observations of the UCFs in ITO nanowires over
a wide range of temperature from below 1 K up to above 10 K now provides us a unique opportunity to verify
these subtle UCF theory predictions.

\begin{figure}
\begin{center}
\includegraphics[scale=0.36]{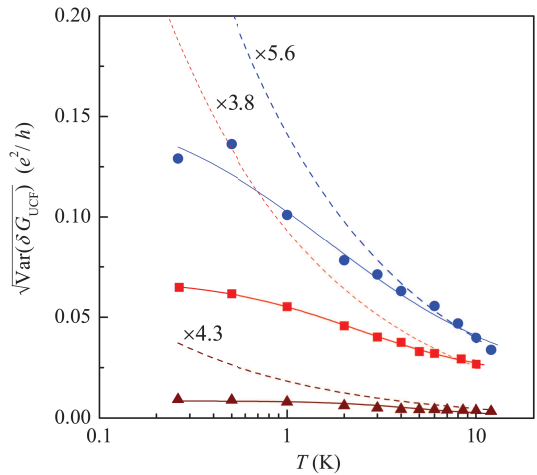}
\caption{Variation of measured $\sqrt{{\langle (\delta G_{\rm UCF})^2 \rangle}}$ (denoted by the square root of the variance $\sqrt{{\rm Var} (\delta G_{\rm UCF})}$ in ordinate) magnitudes with temperature for three ITO nanowires with diameter and length: circles (110 nm and 1.2 $\mu$m), squares (78 nm and 1.4 $\mu$m), and triangles (72 nm, 2.8 $\mu$m). The solid curves drawn through the data points are guides to the eye. The dashed curves are the UCF theory predictions by taking into account both the classical self-averaging and the thermal averaging effects. The thermal averaging effect due to $L_T$ varies approximately as $1/\sqrt{T}$, while the self-averaging effect due to $L_\varphi$ possesses a very weak temperature dependence. The theoretical and experimental values are normalized for 10 K in each nanowire. Note that the theoretical predictions diverge significantly from the experimental results. This figure was reproduced with permission from \cite{YangPY-PRB2012}. Copyright 2012 by the American Physical Society.
\label{UCFs_cutoff}}
\end{center}
\end{figure}

Figure~\ref{UCFs_cutoff} shows a plot of the variation of measured $\sqrt{{\langle (\delta G_{\rm UCF})^2
\rangle}}$ with temperature for three ITO nanowires studied by Yang \etal \cite{YangPY-PRB2012}. Surprisingly,
the theoretical predictions invoking the thermal averaging effect (dashed curves) diverge significantly from
the measured UCF magnitudes (symbols). In figure~\ref{UCFs_cutoff}, the theoretical curves vary approximately
as $1/\sqrt{T}$, while the experiment reveals a much slower temperature dependence. In other words, the
phase-coherent regime in the 1D UCF phenomenon is not cut off by $L_T$, even though the experiment well
satisfied the condition $k_BT > \hbar/\tau_\varphi$ ($L_T < L_\varphi$). The reason why the thermal averaging
effect played no significant role in figure~\ref{UCFs_cutoff} is not understood. The ITO nanowires make
experimentally feasible to reexamine whether any ingredients in the theoretical concepts for thermal averaging
in mesoscopic physics might have been overlooked (overestimated).

In summary, the UCF phenomena manifest rich and subtle quantum-interference properties of a mesoscopic or
nanoscale structure. They provide crucial information about the impurity configuration in a particular sample.
In ITO nanowires, the UCF signals persist up to 20--30 K. For comparison, recall that in conventional
lithographic metal samples, the UCFs (including magnetic-field dependent UCFs and temporal UCFs
\cite{Beutler-PRL1987}) can only be observed at sub-kelvin temperatures \cite{Wagner-PRL2006}. Such pronounced
conductance fluctuations provide valuable opportunities for critical examinations of the underlying UCF
physics \cite{YangPY-PRB2012,Lien-PRB2011}. The presence of marked UCFs suggest that there must exist a large
amount of point defects in artificially synthesized ITO nanostructures, even though the nanowires exhibit a
single crystalline structure under high-resolution transmission electron microscopy studies.\footnote{We note
that it has recently been found that high levels of point defects appear in most artificially grown
single-crystalline nanostructures, including ITO, RuO$_2$ \cite{Lien-PRB2011}, and ZnO \cite{Chiu-nano-2009a}
nanowires.}

\section{Many-body electron transport in granular metals: Inhomogeneous indium tin oxide ultrathin films}

In this section, we discuss the electrical-transport properties of \textit{inhomogeneous} ITO ultrathin films
(average thickness $\approx$\,5--15 nm) which reveal new many-body physical phenomena that are absent in
\textit{homogeneous} disordered systems. These new physical properties, including logarithmic temperature
dependences of both longitudinal electrical conductivity and Hall transport in a wide range of temperature,
have recently been theoretically predicted \cite{Beloborodov-RMP2007, Efetov-PRB2003,Efetov-EPL2003,
Beloborodov-PRL2003,Kharitonov-PRL2007,Kharitonov-PRB2008}, but not yet experimentally tested in detail.

Generally speaking, granular metals are composite materials that are composed of finely dispersed mixtures of
immiscible metal and insulator grains. In many cases, the insulating constituent may form an amorphous matrix
\cite{Abeles-AP1975,Abeles-book1976}. In terms of electrical-transport properties, three distinct regimes can
be achieved in a given granular system, i.e., the metallic, the insulating (dielectric), and the
metal-insulator transition regimes. These three regimes can be conveniently categorized by a quantity called
$G_T$. Here $G_T$ is the average tunneling conductance between neighboring  (metal) grains and is a key
parameter which determines the global electrical properties of a given granular array. $G_T$ can be expressed
in units of $e^2/\hbar$ and written as $G_T = g_T (2e^2/\hbar)$, where $\hbar$ is the Planck constant divided
by $2\pi$, and $g_T$ is a dimensionless average tunneling conductance. The factor 2 arises from the two
allowed spin directions for a tunneling electron. When $g_T > g_T^c$ ($g_T < g_T^c$) the system lies in the
metallic (insulating) regime. A metal-insulator transition occurs at $g_T = g_T^c$. Here $g_T^c = (1/2\pi
\tilde{d}) \ln(E_c/ \tilde\delta)$ is a critical dimensionless tunneling conductance whose value depends on
the dimensionality of the granular array $\tilde{d}$, where $E_c$ is the charging energy, and $\tilde\delta$
is the mean energy level spacing in a grain (references \cite{Beloborodov-RMP2007,Beloborodov-PRL2003}). In
experiments, the magnitude of $g_T^c$ is of order unity or somewhat smaller
\cite{ZhangYJ-PRB2011,Sun-PRB2010}.

Over decades, there has been extensive theoretical and experimental research on the microstructures and
electrical-transport properties of granular systems \cite{Abeles-book1976}. New discoveries have continuously
been made and a good understanding of the physical properties conceptualized. For example, the giant Hall
effect (GHE) has recently been discovered in Cu$_v$(SiO$_2$)$_{1-v}$ \cite{ZhangXX-PRL2001} and
Mo$_v$(SnO$_2$)$_{1-v}$ \cite{WuYN-PRB2011} granular films under the conditions that the grain size $a \ll
L_\varphi$ and the metal volume fraction $v$ is around the quantum percolation threshold $v_q$
\cite{WanC-PRB2002}. The GHE is a novel physical phenomenon which manifests a huge Hall coefficient $R_H$ that
is enhanced by $\sim$\,3 orders of magnitude when $v$ approaches $v_q$ from the metallic side. The GHE is
theoretically explained to arise from the local quantum-interference effect in the presence of rich
microstructures in a metal-insulator composite constituting of nanoscale granules \cite{WanC-PRB2002}. While
the single-particle local quantum interference causes the new GHE, in the following discussion we shall focus
on the many-body electronic transport properties in granular systems.

In the rest of this section, we concentrate on the region with $g_T \gg 1$ or $g_T \gg g_T^c$. The material
systems that we are interested in can thus be termed `granular metals.' In particular, we shall demonstrate
that inhomogeneous ITO ultrathin films are an ideal granular metal system which provides valuable and unique
playgrounds for critically testing the recent theories of granular metals. These new theories of granular
metals are concerned with the many-body electron-electron ($e$-$e$) interaction effect in inhomogeneous
disordered systems. They focus on the electronic conduction properties in the temperature regime above
moderately low temperatures ($T > g_T \tilde\delta /k_B$) where the WL effect is predicted to be comparatively
small or negligible \cite{Beloborodov-RMP2007,Beloborodov-PRB2004b}. In practice, one can explicitly measure
the $e$-$e$ interaction effect by applying a weak perpendicular magnetic field to suppress the
quantum-interference WL effect.

\subsection{Longitudinal electrical conductivity}

For a long time, the electrical-transport properties of granular metals have not been explicitly considered
theoretically. It has widely been taken for granted that the transport properties would be similar to those in
homogeneous disordered metals \cite{Altshuler-EEbook1985}. It was only recently that Efetov, Beloborodov, and
coworkers have investigated the many-body Coulomb $e$-$e$ interaction effect in granular metals. They
\cite{Beloborodov-RMP2007, Efetov-PRB2003,Efetov-EPL2003, Beloborodov-PRL2003} found that the influences of
\emph{e}-\emph{e} interaction on the longitudinal electrical conductivity $\sigma (T)$ and the electronic
density of states $N(E)$ in granular metals are dramatically different from those in homogeneous disordered
metals. In particular, for granular metals with $g_0 \gg g_T$ and $g_T \gg 1$, the {\em intergrain}
\emph{e}-\emph{e} interaction effect causes a correction to $\sigma$ in the temperature range $g_T
\tilde\delta < k_BT < E_c$. Here $g_0 = G_0/(2e^2/\hbar)$, and $G_0$ is the conductance of a single metal
grain. In this temperature interval of practical experimental interest, the total conductivity is given by
\cite{Efetov-PRB2003, Efetov-EPL2003, Beloborodov-PRL2003}
\begin{eqnarray}\label{Eq.(conductivity)}
\sigma & = & \sigma_0 + \delta \sigma \nonumber \\ & = & \sigma_0 \left[ 1 - \frac{1}{2\pi g_T \tilde{d}} \ln \left( \frac{g_TE_c}{k_BT} \right) \right] \,,
\end{eqnarray}
where $\sigma_0 = G_Ta^{2 - \tilde{d}}$ is the tunneling conductivity between neighboring grains in the
absence of Coulomb interaction, and $a$ is the average radius of the metal grain. Note that the correction
term $\delta \sigma$ is negative and possesses a logarithmic temperature dependence. That is, the Coulomb
$e$-$e$ interaction slightly suppresses intergrain electron tunneling conduction, giving rise to $\delta
\sigma /\sigma_0 \propto - 1/g_T$ for $g_T \gg 1$. This $\delta \sigma \propto \ln T$ temperature law is
robust and independent of the array dimensionality $\tilde{d}$. It should also be noted that this correction
term $\delta \sigma$ does not exist in the EEI theory of homogeneous disordered metals
\cite{Altshuler-EEbook1985}.

Soon after the theoretical prediction of equation~(\ref{Eq.(conductivity)}), the electrical-transport
properties of several granular systems were studied, including Pt/C composite nanowires
\cite{Rotkina-PRB2005,Sachser-PRL2011}, B-doped nano-crystalline diamond films \cite{Achatz-PRB2009}, and
granular Cr films \cite{Sun-PRB2010}. The $\delta \sigma \propto \ln T$ temperature law has been confirmed. In
addition, a large suppression in the electronic density of states around the Fermi energy $N(E_F)$ has been
found in studies of the differential conductances of Al/AlO$_x$/Cr tunnel junctions \cite{Sun-PRB2010}, and
thin Pd-ZrO$_2$ granular films \cite{Bakkali-EPL2013}. This last experimental result also qualitatively
confirmed the prediction of the theory of granular metals \cite{Beloborodov-RMP2007,Beloborodov-PRB2004a}.
However, a quantitative comparison is not possible, due to the lack of a theoretical expression for $N(T,V)$
at finite voltages and finite temperatures.

\begin{figure}
\begin{center}
\includegraphics[scale=1.4]{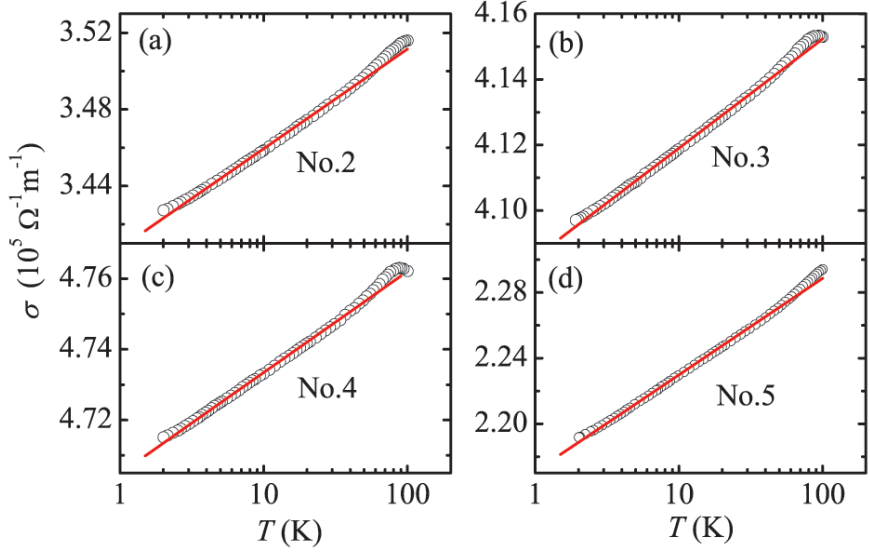}
\caption{Variation of longitudinal electrical conductivity with logarithm of temperature for four inhomogeneous ITO ultrathin films measured in a perpendicular magnetic field of 7 T. The average film thickness (in nm), average grain size (in nm) and fitted $g_T$ value in each film are (a) 9.2, 28 and 13, (b) 11, 34 and 23, (c) 13, 38 and 31, and (d) 7.6, 24 and 7.4. The solid straight lines are least-squares fits to equation~(\ref{Eq.(conductivity)}). This figure was reproduced with permission from \cite{ZhangYJ-PRB2011}. Copyright 2011 by the American Physical Society.} \label{FigZhangPRB4}
\end{center}
\end{figure}

Figure~\ref{FigZhangPRB4} shows the variation of longitudinal electrical conductivity with logarithm of
temperature for four inhomogeneous ITO ultrathin films studied by Zhang \etal \cite{ZhangYJ-PRB2011}. These
films were grown by the RF deposition method onto glass substrates. They were $\approx$ 10$\pm$3 nm thick, and
the average grain sizes were in the range $\approx$\,24--38 nm. Therefore, the samples can be treated as 2D
random granular arrays. (Each sample was nominally covered by one layer of ITO granules.) The conductivities
were measured in a perpendicular magnetic field of 7 T in order to suppress any residual 2D WL effect.
Inspection of figure~\ref{FigZhangPRB4} clearly demonstrates a $\delta \sigma \propto \ln T$ variation over a
wide temperature range from $\sim$\,3 K to $T^\ast$, where  $T^\ast = T^\ast (E_c)$ is the maximum temperature
below which the $\ln T$ law holds. Therefore, the prediction of equation~(\ref{Eq.(conductivity)}) is
confirmed. Quantitatively, from the least-squares fits (the straight solid lines in
figure~\ref{FigZhangPRB4}), values of  the intergrain tunneling conductance $g_T \simeq$ 7--31 were obtained.
Therefore, the theoretical criterion of $g_T \gg 1$ for equation~(\ref{Eq.(conductivity)}) to be valid is
satisfied. We reiterate that the $\delta \sigma \propto \ln T$ temperature law observed in
figure~\ref{FigZhangPRB4} is not due to the more familiar 2D EEI effect which widely appears in homogeneous
disordered systems \cite{ZhangYJ-PRB2011}.

\subsection{Hall transport}

Apart from the longitudinal electrical conductivity, Kharitonov and Efetov
\cite{Kharitonov-PRL2007,Kharitonov-PRB2008} have investigated the influence of Coulomb interaction on the
Hall resistivity, $\rho_{xy}$, by taking the electron dynamics {\em inside} individual grains into account.
They found that there also exists a correction to the Hall resistivity in the wide temperature range $g_T
\tilde\delta \lesssim k_B T \lesssim \min(g_T E_c, \, E_{\rm{Th}})$, where $E_{\rm{Th}} = D_0 \hbar /a^2$ is
the Thouless energy of a grain of radius $a$, $D_0$ is the electron diffusion constant in the grain, and
$\min(g_T E_c, \, E_{\rm{Th}})$ denotes the minimum value of the set $(g_T E_c, \, E_{\rm{Th}})$. The
resulting Hall resistivity is given by \cite{Kharitonov-PRL2007, Kharitonov-PRB2008}
\begin{eqnarray}\label{Eq.(Hall)}
\rho_{xy} (T) & = & \rho_{xy,0} + \delta \rho_{xy} \nonumber \\ & = & \frac{B}{n^\ast e} \left[ 1 + \frac{c_d}{4\pi g_T} \ln \left( \frac{\min(g_T E_c,\,E_{\rm{Th}})}{k_BT} \right) \right] \,,
\end{eqnarray}
where $n^\ast$ is the effective carrier concentration, $c_d$ is a numerical factor of order unity, and
$\rho_{xy,0} = B/(n^\ast e)$ is the Hall resistivity of the granular array in the absence of the Coulomb
$e$-$e$ interaction effect. We point out that the microscopic mechanisms leading to the $\ln T$ temperature
behaviors in equations~(\ref{Eq.(conductivity)}) and (\ref{Eq.(Hall)}) are distinctly different. The
longitudinal conductivity correction $\delta \sigma$ originates from the renormalization of intergrain
tunneling conductance $g_T$, while the Hall resistivity correction $\delta \rho_{xy}$ stems from virtual
electron diffusion inside individual grains \cite{Kharitonov-PRL2007,Kharitonov-PRB2008}.

As mentioned previously, the theoretical predication of equation~(\ref{Eq.(conductivity)}) has been
experimentally tested in a few granular systems. On the contrary, the prediction of equation~(\ref{Eq.(Hall)})
is far more difficult to verify in real material systems. The major reason is due to the fact that the
$\rho_{xy,0}$ magnitude ($\propto 1/n^\ast$) in a granular metal with $g_T \gg 1$ is already small and
difficult to measure. Obviously, the $e$-$e$ interaction induced correction term $\delta \rho_{xy}$ is even
much smaller. Typically, the ratio $\delta \rho_{xy}/\rho_{xy,0} \sim 1/g_T$ is on the order of a few percent
and equation~(\ref{Eq.(Hall)}) is a perturbation theory prediction.

In section 2, we have stressed that the carrier concentration in the ITO material is $\sim$\,2 to 3 orders of
magnitude lower than those in typical metals. Thus, generally speaking, the Hall coefficient, $R_H =
\rho_{xy}/B$, in ITO granular films would be $\sim$\,2 to 3 orders of magnitude larger than those in
conventional granular films made of normal-metal granules. The theoretical predication of
equation~(\ref{Eq.(Hall)}) can hence be experimentally tested by utilizing inhomogeneous ITO ultrathin films.

\begin{figure}
\begin{center}
\includegraphics[scale=1.4]{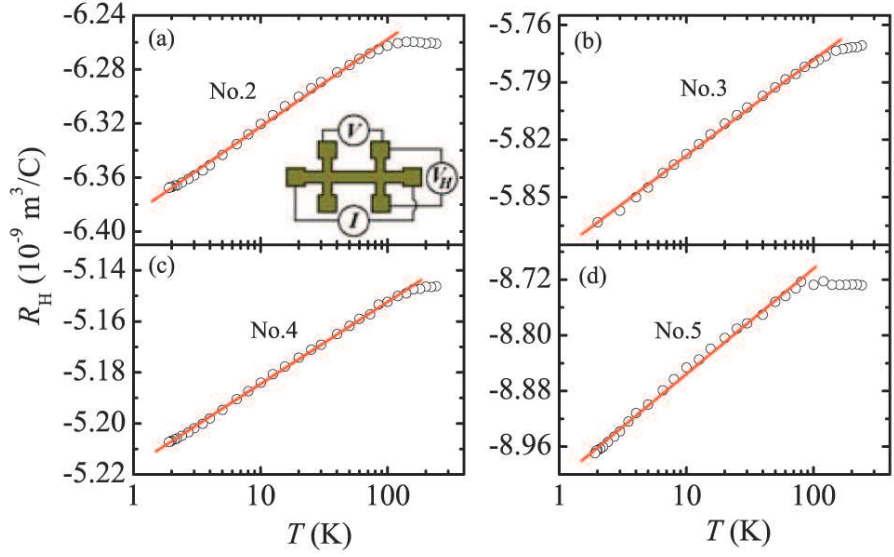}
\caption{Hall coefficient as a function of logarithm of temperature for four inhomogeneous ITO ultrathin films. These films are the same as those shown in figure~\ref{FigZhangPRB4}. The solid straight lines are least-squares fits to equation~(\ref{Eq.(Hall)}). The inset in (a) depicts a schematic for the Hall-bar-shaped sample. This figure was reproduced with permission from \cite{ZhangYJ-PRB2011}. Copyright 2011 by the American Physical Society.} \label{FigZhangPRB3}
\end{center}
\end{figure}

In addition to the observation in figure~\ref{FigZhangPRB4}, Zhang \etal \cite{ZhangYJ-PRB2011} have studied
the Hall transport in inhomogeneous ITO ultrathin films. Figure~\ref{FigZhangPRB3} shows the temperature
dependence of $R_H$ for four samples they have measured. Evidently, one sees a robust $R_H \propto \ln T$
variation over a wide temperature range from $\sim$\,2 K to $T_{\rm max}$, where $T_{\rm max}$ is a
temperature below which the $\ln T$ law holds. The $T_{\rm max}$ value for a given granular array is
determined by the constituent grain parameters $E_c$ and $E_{\rm Th}$ as well as the intergrain tunneling
parameter $g_T$. For those ITO ultrathin films shown in figure~\ref{FigZhangPRB3}, the experimental $T_{\rm
max}$ values varied from $\sim$\,50 to $\sim$\,120 K. Quantitatively, the correction term contributes a small
magnitude of $[R_H(2\,{\rm K}) - R_H (T_{\rm max})] / R_H (T_{\rm max}) \simeq \delta \rho_{xy}(2\,{\rm K})
/\rho_{xy,0} \lesssim$ $5\%$, where $R_H(T_{\rm max}) \simeq 1/(n^\ast e)$ is the Hall coefficient in the
absence of the Coulomb $e$-$e$ interaction effect. The experimental data (symbols) can be well described by
the theoretical predictions (solid straight lines) with satisfactory values of the adjustable parameters.
Thus, the prediction of equation~(\ref{Eq.(Hall)}) is experimentally confirmed for the first time in the
literature.

In summary, the simultaneous experimental observations of $\delta \sigma \propto \ln T$
(figure~\ref{FigZhangPRB4}) and $\delta \rho_{xy} \propto \ln T$ (figure~\ref{FigZhangPRB3}) laws over a wide
range of temperature from liquid-helium temperature up to and above liquid-nitrogen temperature strongly
support the recent theoretical concepts for charge transport in granular metals, i.e.,
equations~(\ref{Eq.(conductivity)}) and (\ref{Eq.(Hall)}), which are formulated under the condition that the
intergrain tunneling conductivity $g_T \gg 1$. We note again that the free-carrier-like and, especially, the
low-$n$ characteristics of the ITO material have made possible a close experimental examination of
equation~(\ref{Eq.(Hall)}). While measurements of $\delta \sigma$ are relatively easy, finding a proper
granular metal with $g_T \gg 1$ to measure the small correction term $\delta \rho_{xy}$ is definitely
nontrivial. The ITO material made into an inhomogeneous ultrathin film form has opened up avenues for
exploring the many-body Coulomb effects in condensed matter physics.

Recently, the thermoelectric power in the presence of granularity and in the limit of $g_T \gg 1$ has been
theoretically calculated \cite{Beloborodov-PRB2009}. It was predicted that the granularity could lead to
substantial improvement in thermodynamic properties and, in particular, the figure of merit of granular
materials could be high. Experimental investigations in this direction would be worthwhile in light of the
development of useful thermoelectric materials. On the other hand, it has recently been reported that the
presence of granularity causes an enhancement of the flicker noise (1/$f$ noise) level in ITO films. This is
ascribed to atomic diffusion along grain boundaries or dynamics of two-level systems near the grain boundaries
\cite{Yeh-APL2013}. Since the 1/$f$ noise could potentially hinder the miniature device performance, it would
be of interest and importance to explore its properties in inhomogeneous ITO ultrathin films.

\section{Conclusion}

Indium tin oxide (ITO) is a very interesting and useful transparent conducting oxide (TCO) material. It is
stable at ambient conditions and can be readily grown into a variety of forms, including polycrystalline thin
and thick films, and single-crystalline nanowires. They can simultaneously have electrical resistivities as
low as $\approx$\,150 $\mu \Omega$\,cm at room temperature and optical transparencies as high as
$\approx$\,90\% transmittance at the visible light frequencies. Apart from their technological issues, the
electronic conduction properties of ITO have rarely been systematically explored as a condensed matter physics
research subject and down to fundamental levels. In this Topical Review, we have focused on metallic ITO
structures. We have shown that the overall electrical resistivity and thermoelectric power can be described by
the Boltzmann transport equation. A linear dependence on temperature of thermoelectric power in a wide range
of temperature eloquently manifests the free-carrier-like energy bandstructure around the Fermi level of this
class of material. At liquid-helium temperatures, marked weak-localization effect and universal conductance
fluctuations emerge. ITO provides a rich playground for studying these quantum interference phenomena in all
three dimensions, which leads to an improved understanding of the underlying physics governing the properties
of mesoscopic and nanoscale structures. Inhomogeneous ITO ultrathin films have opened up unique and valuable
avenues for studying the many-body electron-electron interaction effect in granular metals. These new
theoretical predictions cannot be addressed by employing conventional granular systems.

The objective of this Topical Review is not only to present the charge transport properties of ITO but also to
demonstrate that the ITO material is versatile and powerful for unraveling new physics. Microscopically, the
intrinsic electronic properties that make ITO an appealing technological as well as academic material are the
free-carrier-like energy bandstructure and a low level of carrier concentration. Owing to the inherent
free-carrier-like characteristics, the electronic parameters can be reliably evaluated through the
free-electron model, which in turn facilitate critical tests of a variety of lasting and new theoretical
predictions. A low carrier concentration gives rise to slow electron-phonon relaxation, which manifests the
linear electron diffusion thermoelectric power and also yields a weak electron dephasing rate in the ITO
material. In light of the development and search for useful TCOs, it would be of great interest to investigate
whether the numerous aspects of the novel electronic conduction properties that we have addressed in this
Topical Review might also manifest in other, such as ZnO- and SnO$_2$-based, TCO materials.

\ack{}

The authors thank Yuri Galperin, Andrei Sergeev, and Igor Beloborodov for valuable suggestions and comments,
and David Rees for careful reading of the manuscript. We are grateful to Shao-Pin Chiu, Yi-Fu Chen, Chih-Yuan
Wu, Yao-Wen Hsu, Bo-Tsung Lin, Ping-Yu Yang, and Yu-Jie Zhang for their collaborations at the various stages
of our lasting research on ITO. One of us (JJL) also would like to thank Hsin-Fei Meng for incidentally
igniting his interest in the marvelous electronic conduction properties of the ITO material a decade ago. This
work was supported at NCTU by the Taiwan Ministry of Science and Technology through Grant No. NSC
102-2120-M-009-003 and the MOE ATU Program, and at TJU by the NSF of China through Grant No. 11174216 and the
Research Fund for the Doctoral Program of Higher Education through Grant No. 20120032110065.

\end{document}